# Multi-class within-day dynamic traffic equilibrium with strategic travel time information

Xiaoyu Ma, Xiaozheng He*

Department of Civil and Environmental Engineering, Rensselaer Polytechnic Institute, Troy, NY 12180

**ABSTRACT**

Most research on within-day dynamic traffic equilibrium with information provision *implicitly* considers travel time information, often assuming information to be perfect or imperfect based on travelers' perception error. However, lacking explicit formulation of information limits insightful analysis of information impact on dynamic traffic equilibrium and the potential benefits of leveraging information provision to improve system-level performance. To address this gap, this paper proposes a within-day dynamic traffic equilibrium model that *explicitly* formulates strategic information provision as an endogenous element. The proposed model considers travelers' reactions to the information, creating an interdependent relationship between provided information and traffic dynamics. In this framework, two classes of travelers receive different types of travel time information: one class receives instantaneous travel time reflecting the prevailing traffic conditions, while the other class receives strategic forecasts of travel times, generated by accounting for travelers' reactions to instantaneous information based on strategic thinking from behavioral game theory. The resulting multi-class within-day dynamic equilibrium differs from existing models by explicitly modeling information provision and consideration of information consistency. The inherent dynamics of real-time updated traffic information, traffic conditions, and travelers' choice behaviors are analytically modeled, with the resulting dynamic equilibrium formulated as a fixed-point problem. The theoretical propositions and numerical findings offer rich insights into the impact of information on the traffic network, strategic forecast information penetration, the relationship between the proposed equilibrium and traditional dynamic traffic equilibria, and information accuracy. This research contributes to a deeper understanding of the interplay between information and traffic dynamics, paving the way for more effective traffic management strategies.

* Corresponding author.

E-mail address: hex6@rpi.edu

# 1. Introduction

Advanced Traveler Information Systems (ATIS) play a crucial role in improving surface transportation safety, mobility, and sustainability by providing travelers with essential information such as incident locations, travel times, route guidance, and lane restrictions. Such information, widely disseminated through navigation apps (e.g., Google and Waze), variable message signs, radio, and telephone services (e.g., 511), is poised for significant enhancement with the rapid evolution of big data, machines learning, high performance computing, and emerging vehicular technologies.

As vehicle communication and automation technologies continue to advance, the delivery of travel information is poised to undergo significant enhancements through innovations such as connected vehicle technology. This technology facilitates the seamless exchange of real-time traffic information among vehicles, roadside units, and other road users. In addition, the integration of computation technologies, including cloud and edge computing, enables real-time computations to support smart travel. Consequently, real-time traffic information is expected to play an increasingly pivotal role in the future, to form more intelligent and connected traffic networks (Wang et al., 2021a, Alobeidyeen et al., 2023, Mamdoohi and Miller-Hooks, 2024, Ma and He, 2024). In this context, strategically providing travel information becomes possible and influential, as it not only reflects current traffic conditions but also anticipates travelers' reactions to this information.

How to leverage advanced information provision to enhance traffic network performance requires fundamental research on the impact of traffic information provision. Over decades, research has focused on modeling and exploring the impact of information provision on various aspects of travel behaviors and traffic network performance. Relevant research includes investigating the impacts on four perspectives: (1) travel behaviors on simple networks (e.g., single-bottleneck network and two-parallel-route network), (2) static traffic equilibrium, (3) day-to-day traffic evolution, and (4) within-day traffic dynamics.

This paper contributes to within-day traffic dynamics under strategic information provision, specifically focusing on theoretical equilibrium analysis. In the relevant literature, there are two modeling approaches regarding traffic information, i.e., *implicit* and *explicit*. *Implicit* models do not specify the details of information but rely on underlying assumptions to capture the effect of information. For example, implicit models simply assume that travelers with access to ATIS receive perfect information and follow deterministic user optimal travel behavior (Huang and Lam, 2003, Lee, 2008), or follow stochastic user optimal travel behavior with less uncertainty and perception error (Li and Su, 2005); while travelers without access to ATIS are assumed to follow stochastic user optimal travel behavior with larger uncertainty and perception error (Huang and Lam, 2003, Lee, 2008, Li and Su, 2005). This type of research contributes to the consideration of traffic information in dynamic traffic assignment in an analytical way. However, since the information is considered implicitly, it does not allow for the analysis of the effect of specific traffic information, insights into information design, or travelers' reactions to the information.

*Explicit* models incorporate the modeling of the specific content of traffic information. The explicit information can be quantitative or descriptive traffic conditions. With this explicit approach, a crucial concept is *information consistency*, which is satisfied when the traffic forecasts on which the information is based are verified after drivers react to the information (Bottom, 2000). This concept stems from the following real-world phenomenon: When information based on whatever traffic condition forecasts are disseminated, travelers' reactions to the information may invalidate the forecasts and render the information irrelevant and self-defeating. For instance, if forecast information predicts impending congestion in one corridor, travelers may choose to switch to an alternative corridor, leaving the original one relatively uncongested (Bottom, 2000). This phenomenon can lead to potential issues such as instable traffic, compromised information accuracy, and reduced trustworthiness. Hence, information consistency becomes essential in the modeling of traffic equilibrium with explicit information provision.

This paper *explicitly* models the traffic information for within-day traffic dynamic equilibrium modeling and analysis. In the literature, explicit traffic information provision has been primarily studied through simulation, with a notable lack of analytical models and theoretical analysis. Among the relevant studies, only two (Bottom, 2000, Hoang et al., 2023) have explicitly modeled the provided information. However, their focus differs significantly from the present research, which will be discussed in Section 2. To the best of our knowledge, this paper represents the first attempt to theoretically model and analyze the dynamic traffic equilibrium under *explicit* traffic information provision. The explicit modeling of travel time information, along with the incorporation of information consistency and travelers' reactions to information, highlights the significant impact of strategic information provision. This approach sets our research apart from existing studies and underscores its significance in advancing this field.

In this paper, traffic information takes the form of travel times. In this context, information consistency does not necessitate the traffic forecasts to precisely match the resulting traffic after travelers' reactions to information, because such a requirement implies the strong assumption of perfect information, i.e., 100% accurate travel time forecasts. Namely, if the traffic forecasts perfectly account for travelers' reactions to information, it will create an infinite loop in generating the forecast information, requiring continuous forecasting to capture how travelers react to forecast information and leading to an unrealistic and endless cycle. Consequently, the provided information in this paper is realistically considered *imperfect*. Two traveler classes are defined: one receiving instantaneous travel time information reflecting prevailing traffic conditions, and the other receiving strategic forecasts of travel times that consider travelers' reactions to instantaneous information. Both types of information are updated in real-time as traffic dynamics evolve.

In our proposed model, information consistency forms a closed loop involving three elements: traffic condition, travel time information, and travel behaviors including simultaneous departure time and route choices. As discussed in (Bottom, 2000), a model addressing information consistency with explicit information encompasses three mappings: (1) information generation mapping, from traffic condition to traffic information; (2) travel choice mapping, from traffic information to travelers' choice behaviors; (3) dynamic network loading mapping, from travel behaviors to traffic condition. In our research, information consistency is achieved when these three elements compose a closed loop through the three mappings. That is, the traffic condition on which the information generation is based is consistent with the traffic condition resulting from travelers' reactions to information. Indeed, the closed loop can commence from any of the three elements, progressing through a series of mappings, ultimately returning to the same starting element. Details about the three mappings modeled in this research will be presented in Section 4.

This research contributes to the modeling and analysis of explicit strategic traffic information provision and its resulting dynamic traffic equilibrium. It pioneers the explicit modeling of dynamic information provision, capturing it as an endogenous element within the dynamic traffic equilibrium—an approach that has never been explored in the literature. By incorporating strategic information provision, the proposed dynamic traffic equilibrium model obtains broader flexibility to represent real-world complexity compared to traditional models, with the traditional stochastic dynamic user equilibrium as its special case under certain conditions. This research serves as the theoretical foundation for understanding the impacts of strategic information provision, particularly in the era of connected vehicle systems, thereby advancing the field and offering new insights into effective traffic management strategies.

In summary, this research makes the following specific contributions: (1) It advances the modeling and analysis of dynamic traffic equilibrium by *explicitly* formulating travel time information provision; (2) The proposed model captures information consistency and travelers' reactions to information without relying on strong assumptions of information perfection; (3) This research introduces a closed loop that encompasses mappings among traffic condition, travel time information, and travel behaviors, making information provision an endogenous element in the dynamic traffic equilibrium. This sets it apart from traditional dynamic traffic equilibrium models; (4) The complexities introduced by the dynamic nature of the real-time information provision, evolving traffic condition, and travel behaviors are effectively addressed through rigorous modeling and analysis; (5) The theoretical propositions and numerical findings provide rich insights into the impact of information on travelers and the traffic network, the penetration of forecast information, the relationship with traditional dynamic traffic equilibria, and information accuracy.

The remainder of the paper is organized as follows. Section 2 presents a review of related work. Section 3 presents the problem statement, including the modeling framework, definitions of terminologies, and an illustrative example. Section 4 defines and formulates the proposed dynamic traffic equilibrium and the corresponding fixed-point problem. Section 5 presents the theoretical properties of the proposed dynamic traffic equilibrium, including the equivalence to the fixed-point problem, existence, relationship with the existing dynamic equilibrium models, and information accuracy analysis. Section 6 provides a solution algorithm used for solving the equilibrium. Section 7 conducts numerical examples to demonstrate the proposed work and draw insights from the results. Section 8 concludes this research.

## 2. Related Work

From a broader point of view, the research relevant to the impacts of travel information on traffic networks and travel behaviors can be categorized into four main perspectives: (1) *travel behaviors in simple networks* (e.g., single-bottleneck network and two-parallel-route network) (Yu et al., 2021, Wang et al., 2021b, Tavafoghi and Teneketzis, 2017, Liu and Yang, 2021, Lindsey et al., 2014, Khan and Amin, 2018, Han et al., 2021, Arnott et al., 1991, Ahmed et al., 2015), (2) *static traffic equilibrium* (Yang, 1998, Xie and Liu, 2022, Nakayama, 2016, Li et al., 2017, Jiang et al., 2020, Huang and Li, 2007, Gao and Huang, 2012, Ettema and Timmermans, 2006, Acemoglu et al., 2018), (3) *day-to-day traffic evolution* (Zhao et al., 2019, Yu et al., 2020, Ye et al., 2021, Yang and Jayakrishnan, 2013, Xu et al., 2014, Liu et al., 2017, Li et al., 2018, Delle Site, 2018, Bifulco et al., 2016), and (4) *within-day traffic dynamics* (Xiong et al., 2017, Wahle et al., 2002, Lu and Yang, 2009, Li and Su, 2005, Levinson, 2003, Lee, 2008, Kucharski and Gentile, 2019, Huang and Lam, 2003, Hoang et al., 2023, Bottom, 2000).

This section presents the review of the fourth category that is the most relevant to the proposed research, i.e., the within-day traffic dynamics under information provision. In this category, two main branches exist, including theoretical equilibrium analysis (i.e., dynamic traffic assignment) and simulation studies.

The proposed research is under the branch of theoretical equilibrium modeling and analysis. In the literature of this branch, most studies do not explicitly model the provided information but have simple travel behavior assumptions based on the information type. For example, Huang and Lam (2003) propose a multi-class dynamic traffic equilibrium model. In their model, one traveler class with access to ATIS is assumed to have perfect information, following deterministic user optimal behavior. In contrast, the other class without ATIS access is assumed to have incomplete information, following stochastic user optimal behavior. Similarly, Li and Su (2005) also consider a multi-class dynamic traffic equilibrium and assume that travelers with access to ATIS have less uncertainty in travel cost while travelers without access to ATIS have higher uncertainty in travel costs and both classes follow stochastic user optimal behavior. Additionally, Lee (2008) consider a three-class dynamic traffic equilibrium with both mode choice and route choice, where travelers with full access to ATIS follow user optimal behavior, and travelers with limited access to ATIS follow stochastic optimal behavior with different levels of uncertainty based on the degree of ATIS access. Note that these studies do not investigate the effect of specific explicit traffic information, do not consider the reaction of travelers to the information provided, and information consistency is not applicable; therefore, clearly differing in research interests from the present paper.

While two research studies (Bottom, 2000, Hoang et al., 2023) consider explicit information provision, their focus emphasizes different research settings and problems compared with this paper. Bottom's dissertation Bottom (2000) focuses on providing a general framework for generating and computing route guidance information that ensures information consistency, instead of mathematically formulating a specific dynamic traffic equilibrium and conducting theoretical analysis. The dissertation intends to provide a simulator for numerical computation of traffic information and validation of different algorithms to address the stochasticity in enroute travel behaviors and traffic conditions. Hoang et al. (2023) propose a framework of two-class dynamic traffic assignment with instantaneous traffic information presented as a scenario set characterizing traffic demand and capacity. They assume one class of travelers to be selfish, following user optimal behavior, while the other class is fully cooperative, adhering to system optimal behavior.

The other branch of research (Kucharski and Gentile, 2019, Levinson, 2003, Lu and Yang, 2009, Wahle et al., 2002, Xiong et al., 2017) conducts micro simulations or numerical experiments to explore information's impact on within-day traffic dynamics, especially on travelers' enroute behaviors. Though implicit information provision has been considered in this branch, they are simulation studies instead of analytic models or theoretical analysis. Since this branch deviates from the paper's research scope, details are omitted for brevity.

As reviewed, within the domain of theoretical modeling and analysis of within-day dynamic traffic equilibrium with information provision, this paper stands out as the pioneering work that explicitly models travel time information provision, while accounting for information consistency and travelers' reactions. By contrast, previous studies that have employed explicit traffic information provision are primarily through simulation, with a lack of analytical models and theoretical analysis. The proposed dynamic traffic equilibrium framework distinguishes itself from existing models by embedding the travel time information as an endogenous element, laying the foundation for theoretical analysis of the impact of traffic information on within-day traffic dynamics. Its novelty is further demonstrated by an illustrative example in the following section.

## 3. Problem Statement

This paper aims to develop a dynamic traffic equilibrium model under explicit information provision, while ensuring information consistency. The model incorporates two types of information including instantaneous travel time and forecast travel time, which are disseminated to two classes of travelers. Travelers make simultaneous path-and-departure-time (SPDT) choices based on the received information. Both types of information are updated in real-time, reflecting the evolving traffic conditions influenced by travelers' choices. Travelers do not switch paths once they depart, i.e., no enroute behavior considered.

The proposed framework establishes a closed loop involving traffic condition, travel time information, and travelers' choice behaviors, ensuring information consistency. This closed loop is inherently dynamic, with traffic conditions dynamically shaped by travelers' SPDT choices over time; and the evolving traffic condition, in turn, generates real-time updated information that influences subsequent SPDT choices of travelers. Figure 1 shows the closed loop framework. The dynamic traffic equilibrium capturing this dynamic closed loop will be modeled and formulated as a fixed-point problem in Section 4, including the modeling of the three mappings.

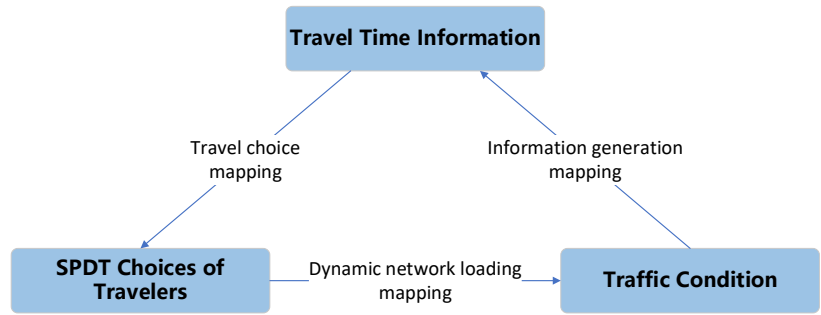

Figure 1. Modeling framework

*3.1 Definitions*

To avoid any confusion, the following provides explanations for several terminologies used in this paper.

*Instantaneous information*. The instantaneous travel time information reflects the prevailing traffic condition at the current time. In practice, it can be obtained through detectors, mobile devices, or roadside units. The instantaneous path travel time is the summation of the current travel time of each link that the path involves, without factoring possible traffic dynamics that could bring changes in the travel times of subsequent links along the path during the trip. At each discrete time interval, travelers who receive instantaneous information will be provided with the instantaneous path travel times for the current time of departure only. Instantaneous information is updated in real-time in relation to the evolving traffic dynamics.

*Forecast information*. The forecast travel time information is obtained by forecast path departures. In this paper, it is obtained by factoring the travelers' reactions to instantaneous information, which refers to a mechanism called "strategic thinking". Section 4 will present the details. The forecast path travel time intends to provide a forecast of the travel time that is going to take for traveling on the path, but it is unnecessary to be perfect. At each discrete time interval, travelers who receive forecast information will be provided with a series of forecast path travel times for the departure times from the current time onwards, for instance, for the next 10 minutes of departure.

*Realized travel time*. The realized travel time is the actual experienced travel time after travelers complete their trip. It may differ from the provided information, indicating imperfect information provision.

*Tentative travel choices*. At each time interval, both classes of travelers who have not departed make tentative SPDT choices based on their received information. Those who chose to depart at the current time interval will indeed depart, while those tentatively chose to depart later will wait. When the next time interval starts, those who have not departed will receive newly updated information and based on this information, they will make new tentative SPDT choices. This process continues until they depart.

*Realized travel choices / path departures.* As mentioned above, travelers make tentative SPDT choices at each time interval based on the latest received information. Consequently, only the choices made by those who decided to depart "right now", i.e., at the current time interval, will be realized as actual departures, which are referred to as realized travel choices / path departures. The tentative SPDT choices made by those who wait will not be realized, as new tentative SPDT choices will be generated in the next time interval. In essence, not all tentative SPDT choices will be realized as actual departures.

*3.2 Illustrative example*

This section presents an illustrative example of a simple network to enhance the understanding of the proposed framework for the problem of interest. Figure 2 shows a network with three links connecting two origin-destination (OD) pairs, A to C and B to C, denoted as OD pair AC and BC, respectively. OD pair AC is connected by two paths: Path 1 involves Links 1 and 3 and Path 2 involves Links 2 and 3. OD pair BC is connected by a single path, denoted as Path 3, which involves link 3. Consider a time horizon containing three time intervals for departure. There are 12 travelers for each OD pair, with 6 in each class. Table 1 provides an overview of the information provided, tentative SPDT choices, as well as realized SPDT choices in the three departure time intervals. In the table, the matrices or vectors are composed of rows representing paths and columns representing time intervals. The three rows from top to bottom of the matrices correspond to Path 1, Path 2, and Path 3, respectively. The numerical values are unit-free, for illustrative purposes.

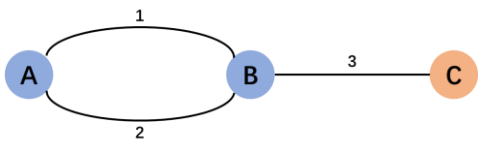

Figure 2. Illustrative network

At each time interval, the vectors, *a*, *b*, and *c*, are the instantaneous travel time information generated based on certain traffic condition at that moment, respectively. These vectors only have one column, because they provide the instantaneous path travel times for the current time interval only. Differently, the forecast information, matrices *d*, *e*, and *f*, can have multiple columns, since they provide forecast path travel times for the current time interval onwards. For example, the columns from left to right of the matrix *d* represent the forecast path travel times for departure at the $1^{st}$, $2^{nd}$, and $3^{rd}$ time intervals, respectively; and the columns from left to right of the matrix *e* represent the forecast path travel times for departure at the $2^{nd}$ and $3^{rd}$ time intervals, respectively. Since in this example, we only have the finite time horizon of interest containing three time intervals, we end up with a one column vector *f* for the forecast information at the $3^{rd}$ time interval. In practice, the forecast could be rolling with a certain length of time window.

Travelers make tentative SPDT choices based on the received information, along with other potential factors such as early/late arrival penalty, at each time interval. In matrices *g – l*, tentative choices are represented by path departures (number of travelers). For example, the first row of the matrix $g$ represents that there is one traveler choosing to depart on Path 1 at the 1st time interval, one traveler tentatively chooses to depart on Path 1 at the 2nd time interval, and another one traveler tentatively chooses to depart on Path 1 at the 3rd time interval. When the 2nd time interval starts, those who have not departed make new tentative choices as they receive new information. Similarly, as seen in the forecast information, the number of columns reduces as time evolves. The bold values in matrices $g - l$ are realized path departures, and compose the matrices $m$ and $n$, for the two classes, respectively. Matrix $o$ is the realized path departures for the whole group of travelers, which is the sum of matrices $m$ and $n$.

The closed loop in Figure 1 forms when the actual traffic condition, mapped from the realized path departures (matrix $o$), is consistent with the traffic condition that the provided information is originally generated from through the information generation mapping, i.e., when information consistency is satisfied. Note that this simple illustrative example aims to illustrate the basic logic of the dynamic framework, and the numbers are provided without utilizing any well-defined mappings in the closed loop. In the following section, all the components and mappings will be modeled, which characterize the dynamic equilibrium.

Table 1. Illustrative example

|  | 1st time interval | 2nd time interval | 3rd time interval | Realized path departures (by class) | Realized path departures |
|---|---|---|---|---|---|
| Instantaneous information | $a. \begin{bmatrix} 3 \\ 3 \\ 1 \end{bmatrix}$ | $b. \begin{bmatrix} 6 \\ 5 \\ 2 \end{bmatrix}$ | $c. \begin{bmatrix} 6 \\ 4 \\ 2 \end{bmatrix}$ |  |  |
| Forecast information | $d. \begin{bmatrix} 3 & 4 & 5 \\ 3 & 4 & 3 \\ 2 & 1 & 1 \end{bmatrix}$ | $e. \begin{bmatrix} 5 & 5 \\ 5 & 4 \\ 2 & 1 \end{bmatrix}$ | $f. \begin{bmatrix} 6 \\ 5 \\ 2 \end{bmatrix}$ |  |  |
| Tentative choices – 1st class | $g. \begin{bmatrix} \mathbf{1} & 1 & 1 \\ \mathbf{1} & 1 & 1 \\ \mathbf{2} & 3 & 1 \end{bmatrix}$ | $h. \begin{bmatrix} \mathbf{0} & 0 \\ \mathbf{3} & 1 \\ \mathbf{3} & 1 \end{bmatrix}$ | $i. \begin{bmatrix} \mathbf{0} \\ \mathbf{1} \\ \mathbf{1} \end{bmatrix}$ | $m. \begin{bmatrix} 1 & 0 & 0 \\ 1 & 3 & 1 \\ 2 & 3 & 1 \end{bmatrix}$ | $o. \begin{bmatrix} 3 & 0 & 0 \\ 3 & 4 & 2 \\ 4 & 5 & 3 \end{bmatrix}$ |
| Tentative choices – 2nd class | $j. \begin{bmatrix} \mathbf{2} & 0 & 0 \\ \mathbf{2} & 1 & 1 \\ \mathbf{2} & 3 & 1 \end{bmatrix}$ | $k. \begin{bmatrix} \mathbf{0} & 0 \\ \mathbf{1} & 1 \\ \mathbf{2} & 2 \end{bmatrix}$ | $l. \begin{bmatrix} \mathbf{0} \\ \mathbf{1} \\ \mathbf{2} \end{bmatrix}$ | $n. \begin{bmatrix} 2 & 0 & 0 \\ 2 & 1 & 1 \\ 2 & 2 & 2 \end{bmatrix}$ |  |

*1st class: travelers receiving instantaneous information; 2nd class: travelers receiving forecast information

Note that in traditional dynamic equilibrium models with implicit information, there is no analytical modeling of information. In particular, traveler's behavior is usually modeled as deterministic user optimal behavior when perfect information is provided, i.e., travelers are pursuing shortest realized travel time; or, it is assumed that travelers follow certain stochastic user optimal behavior when imperfect information (conceptually without analytical modeling) is provided. Differently, in this paper, the information (matrices a-f) is explicitly and analytically modeled, and is generated by capturing potential reactions from travelers. The information is provided dynamically given the evolving traffic dynamics, well capturing reality. This modeling framework makes the information a dynamic and endogenous element intercorrelating with traffic dynamics, which, therefore, provides the foundation for further theoretical analysis of the impact of traffic information on within-day traffic dynamics.

## 4. Model

Consider a traffic network $G(N, L)$ consisting of a set of nodes, $N$, and a set of directed links, $L$. Let $W$ be the set of origin-destination (OD) pairs and $\mathcal{P}_w$ be the set of paths connecting OD pair $w \in W$. Denote $\Delta = (\delta_{a,p})$ as the link-path incidence matrix, where $\delta_{a,p} = 1$ if path $p$ uses link $a$ and 0 otherwise. Assume that the travel demand in the network is fixed and denoted by a vector $\mathbf{d} = (d_w)^T$, where $d_w = d_{I,w} + d_{F,w}$ represents the travel demand between OD pair $w \in W$ with $d_{I,w}$ and $d_{F,w}$ as the demands receiving instantaneous information and forecast information, respectively. The time horizon of interest is discretized into $T$ time intervals. Denote $\mathcal{T} = \{t_1, t_2, t_3, \ldots, t_T\}$ as the set containing all the discretized time intervals and $\mathcal{T}_{\tilde{t}} = \{t | t \geq \tilde{t}, t \in \mathcal{T}\}$ as the set containing the time intervals from $\tilde{t}$ onward to $t_T$. Denote $h_{I,p}(t)$ and $h_{F,p}(t)$ as the number of departures of the two classes on path $p$ at time $t \in \mathcal{T}$, respectively. In the rest of the paper, we omit the $(t)$ in the notations of path departures for simplicity unless the departure time needs to be specified. Let vector $\mathbf{h} = (h_p)^T$ with $h_p = h_{I,p} + h_{F,p}$ being the total departures on path $p$. Denote vectors $\mathbf{h}_I = (h_{I,p})^T$ and $\mathbf{h}_F = (h_{F,p})^T$. Denote $\phi_{I,\tilde{t}}(p)$ as the instantaneous travel time information for path $p \in \mathcal{P}$ provided at time $\tilde{t}$, and let vector $\mathbf{\Phi}_{I,\tilde{t}} = [\phi_{I,\tilde{t}}(p), p \in \mathcal{P}]$. Denote $\phi_{F,\tilde{t}}(p,t)$ as the forecast travel time information provided at time $\tilde{t}$ for the choice pair of path $p \in \mathcal{P}$ and departure time $t \in \mathcal{T}_{\tilde{t}}$. Let matrix $\mathbf{\Phi}_{F,\tilde{t}} = [\phi_{F,\tilde{t}}(p,t), p \in \mathcal{P}, t \in \mathcal{T}_{\tilde{t}}]$.

The sets of feasible path departure vectors are defined as follows.

$$\Lambda_I = \left\{ \mathbf{h}_I \geq 0 : \sum_{p \in \mathcal{P}_w} \sum_{t \in \mathcal{T}} h_{I,p}(t) = d_{I,w} \quad \forall w \in W \right\}$$

$$\Lambda_F = \left\{ \mathbf{h}_F \geq 0 : \sum_{p \in \mathcal{P}_w} \sum_{t \in \mathcal{T}} h_{F,p}(t) = d_{F,w} \quad \forall w \in W \right\}$$

where $\Lambda_I$ and $\Lambda_F$ are the feasible sets of path departure vectors for the two traveler classes, respectively. Other notations will be explained when first presented.

In the following subsections, the Information Generation Mapping and the Travel Choice Mapping in the closed loop shown in Figure 1 will be presented, followed by the definition of the dynamic traffic equilibrium and the formulation of the fixed-point problem. Since the modeling of Dynamic Network Loading (DNL) is a distinct research topic and not the focus of this paper, we directly adopt the one proposed in Han et al. (2019) with a minor adaptation. Note that other DNL models can also be applied within the proposed modeling framework. In Appendix A, we have presented a general description about the DNL model used in numerical experiments for completeness.

*4.1 Information generation*

This subsection presents the information generation mapping, which maps from the traffic condition to travel information. Here, the traffic condition is represented by link/path travel times on the traffic network. As mentioned before, two types of information are considered, i.e., instantaneous and forecast information. Recall that vector $\mathbf{\Phi}_{I,\tilde{t}} = [\phi_{I,\tilde{t}}(p), p \in \mathcal{P}]$ collects the instantaneous travel times for all paths $p \in \mathcal{P}$ at time $\tilde{t}$ and matrix $\mathbf{\Phi}_{F,\tilde{t}} = [\phi_{F,\tilde{t}}(p,t), p \in \mathcal{P}, t \in \mathcal{T}_{\tilde{t}}]$ collects the forecast travel times at time $\tilde{t}$ for departure time $t \in \mathcal{T}_{\tilde{t}}$ on all paths $p \in \mathcal{P}$. Each row of $\mathbf{\Phi}_{F,\tilde{t}}$ represents different paths $p \in \mathcal{P}$ and each column of $\mathbf{\Phi}_{F,\tilde{t}}$ represents different time intervals $t \in \mathcal{T}_{\tilde{t}}$. Note that $\mathbf{\Phi}_{I,\tilde{t}}$ is a vector while $\mathbf{\Phi}_{F,\tilde{t}}$ is a matrix since instantaneous information is provided for the current time only; while the forecast information covers all time intervals in $\mathcal{T}_{\tilde{t}}$, i.e., from time $\tilde{t}$ onward.

The instantaneous travel time information reflects the instantaneous traffic condition, which is directly obtained by a DNL procedure with the input path departure $\mathbf{h}$. Therefore, the instantaneous travel time information is directly generated from the current traffic condition with no additional mapping needed. For convenience, we will write the instantaneous travel time information at time $\tilde{t}$ as $\mathbf{\Phi}_{I,\tilde{t}}(\mathbf{h})$ to represent its relationship with path departure $\mathbf{h}$. In the remaining of this subsection, we will focus on the generation of the forecast information.

The forecast travel time information is generated with the consideration of travelers' reactions to instantaneous information. Specifically, at each time interval, forecast path departures considering the travelers' reaction will be first generated and then from those forecast path departures the forecast travel times will be then generated. The forecast travel time intends to provide a forecast of the realized travel time that is going to take, but it is unnecessary to be perfect. Essentially, the way of generating forecast information is based on the concept of *strategic thinking* in behavioral game theory (Camerer, 2003). Concisely, strategic thinking means that agents make decisions based on a prediction of what others might do (Camerer, 2003, He and Peeta, 2016). Following such a logic, the forecast information is generated based on a prediction of what others might do assuming they behave based on instantaneous information. This prediction can be conducted in real time by an information provider with advanced computation technologies.

The generation of forecast information needs the forecast path departures as input. With forecast path departures, the forecast information can be computed by a DNL procedure. In the following we present the generation of forecast path departures.

At $\tilde{t} \in \mathcal{T}$, the forecast path departures for time $\tilde{t}$ and onward ($t \geq \tilde{t}$), denoted as $\bar{\mathbf{h}}_{\tilde{t}}$, is computed by

$$\bar{\mathbf{h}}_{\tilde{t}} = \mathrm{M}\big(\mathbf{\Phi}_{I,\tilde{t}}(\mathbf{h}), \mathbf{d}_{\tilde{t}}\big), \qquad \mathbf{d}_{\tilde{t}} = \left\{ (d_{\tilde{t},w}) \,\middle|\, d_{\tilde{t},w} = d_w - \sum_{t=1}^{t=\tilde{t}-1} \mathbf{h}_{p \in \mathcal{P}_w}(t) \right\}. \tag{1}$$

where M is the mapping from information to tentative travel choices (path departures), which will be presented in Section 4.2 in detail; $\mathbf{d}_{\tilde{t}}$ denotes the travel demand at time $\tilde{t}$ (those have not departed). Indeed, $\bar{\mathbf{h}}_{\tilde{t}}$ is the forecast path departures predicted at time $\tilde{t}$ for the future departure time intervals, with the input of $\boldsymbol{\Phi}_{I,\tilde{t}}$, i.e., considering travelers' reactions to instantaneous information.

The forecast information can then be obtained by the DNL procedure with input $\bar{\mathbf{h}}_{\tilde{t}}$. Here, in practice, we can ignore the path departures that happened before time $\tilde{t}$, and directly conduct travel time prediction based on $\bar{\mathbf{h}}_{\tilde{t}}$ and the realized traffic condition. However, in the modeling, we will need to include the path departures before time $\tilde{t}$ as well for the input of DNL when doing travel time prediction since the past path departures will affect the computation of travel times for the subsequent future time intervals. That is, the forecast formation at time $\tilde{t}$ is generated with input of both the realized path departures before time $\tilde{t}$ and the forecast path departures for $t \geq \tilde{t}$. Therefore, we have the following formulation of the forecast information.

$$\boldsymbol{\Phi}_{F,\tilde{t}} = \left[\phi_{F,\tilde{t}}(p,t), p \in \mathcal{P}, t \in \mathcal{T}_{\tilde{t}}\right] \tag{2}$$

with

$$\phi_{F,\tilde{t}}(p,t) = DNL(\hbar_{\tilde{t}})(p,t), \qquad \hbar_{\tilde{t}}(t) = \begin{cases} \mathbf{h}(t) & if\ t < \tilde{t} \\ \bar{\mathbf{h}}_{\tilde{t}}(t) & if\ t \geq \tilde{t} \end{cases} \tag{3}$$

where $DNL$ represents a mapping that outputs the forecast path travel times with the input forecast path departures. This formulation again suggests the forecast travel time information is imperfect since it is generated from path departure predictions without knowing the realized path departures in the future.

Note that, as seen in Eq. (3), in the computation of forecast information, the DNL procedure is used at each time of forecasting at $\tilde{t} \in \mathcal{T}$. This adds complexity in computation and theoretical analysis.

Figure 3 shows the workflow of information generation mapping. Given the path departures, the instantaneous travel times can be directly obtained through the DNL procedure. Subsequently, using the instantaneous travel time information, the forecast path departures are obtained at each time interval, based on which the forecast travel time information is generated for each time interval using Eqs. (1)-(3). The figure also clearly shows the difference between the dimension of $\boldsymbol{\Phi}_{I,\tilde{t}}$ and $\boldsymbol{\Phi}_{F,\tilde{t}}$ at each time $\tilde{t} \in \mathcal{T}$, as introduced in the information provision mechanism. Note that, in theoretical modeling, the time horizon must be finite and fixed with a predefined number of time intervals. Therefore, the column dimension of $\boldsymbol{\Phi}_{F,\tilde{t}}$ reduces as time evolves, as any time $t > t_T$ is out of interest.

Also, note that this paper aims to theoretically formulate and analyze the dynamic traffic equilibrium with the proposed information provision. Therefore, other social issues such as fairness, information pricing, information compliance are out of the scope of this paper and could be further discussed in future work. In this paper, it is assumed that the travelers make SPDT choices based on the information they receive without switching class or denying the information. The different types of information can be provided by different information providers.

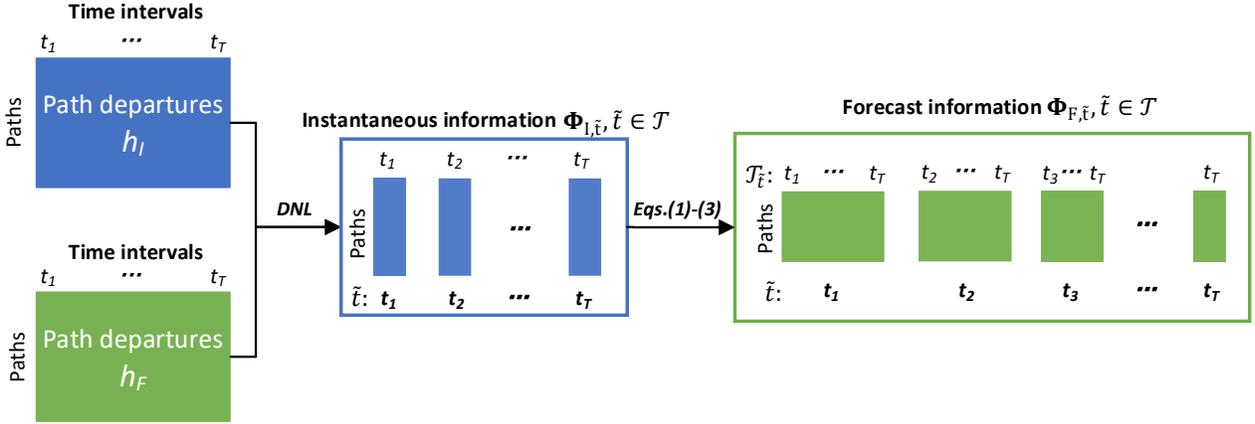

Figure 3. Information generation mapping

*4.2 Travel choices*

This subsection presents the travel choice mapping that maps from the provided information to travelers' SPDT choices. As presented in the illustrative example, travelers make tentative SPDT choices at each time interval until they depart. In the following, we first present how the travelers make tentative SPDT choices based on received information and then convert the tentative choices into realized path departures.

The modeling of the travelers' behaviors can vary based on the different behavioral assumptions. In this paper, we use random utility theory and discrete choice models as the foundation for modeling travelers' tentative choice behaviors. Note that specific behavioral models are not the research focus of this paper, and the proposed framework remains applicable if other behavioral models are adopted. Although the random utility theory and discrete choice models are well-developed, we still present relevant details in the following for completeness and clarity of the proposed work, as well as for their application in numerical experiments.

Based on random utility theory, travelers perceive certain levels of disutility when facing choices. For SPDT choices, the disutility is assumed to be randomly related to travel time and early/late arrival penalty, with a random error term capturing any unobserved attributes. At each time interval $\tilde{t} \in \mathcal{T}$, travelers perceive disutility for each of the available SPDT choice they are facing, i.e., the SPDT choice of any path in the corresponding OD pair and any departure time in $\mathcal{T}_{\tilde{t}}$. Recall $\mathcal{T}_{\tilde{t}} = \{t | t \geq \tilde{t}, t \in \mathcal{T}\}$. Let matrices $\widehat{\mathbf{\Psi}}_{I,\tilde{t}}$ and $\widehat{\mathbf{\Psi}}_{F,\tilde{t}}$ represent the perceived disutility of all choices faced by the two classes of travelers receiving instantaneous and forecast information, respectively, at time $\tilde{t}$. The rows of $\widehat{\mathbf{\Psi}}_{I,\tilde{t}}$ and $\widehat{\mathbf{\Psi}}_{F,\tilde{t}}$ represent paths $p \in \mathcal{P}$ and columns represent time intervals $t \in \mathcal{T}_{\tilde{t}}$. Denote the individual element in matrices $\widehat{\mathbf{\Psi}}_{I,\tilde{t}}$ and $\widehat{\mathbf{\Psi}}_{F,\tilde{t}}$ as $\widehat{\Psi}_{I,\tilde{t}}(p,t)$ and $\widehat{\Psi}_{F,\tilde{t}}(p,t)$, respectively, representing the perceived disutility associated with the SPDT choice of path $p$ and departure time $t$.

At $\tilde{t} \in \mathcal{T}$, the perceived disutility with respect to the SPDT choice of departure time $t \in \mathcal{T}_{\tilde{t}}$ and path $p \in$

$\mathcal{P}$ is defined as

$$\widehat{\Psi}_{I,\tilde{t}}(p,t) = V(\phi_{I,\tilde{t}}(p),t) + \varepsilon_{I,w}, \qquad p \in \mathcal{P}_w, t \in \mathcal{T}_{\tilde{t}}, \tilde{t} \in \mathcal{T}, w \in W \tag{4}$$

$$\widehat{\Psi}_{F,\tilde{t}}(p,t) = V(\phi_{F,\tilde{t}}(p,t),t) + \varepsilon_{F,w}, \qquad p \in \mathcal{P}_w, t \in \mathcal{T}_{\tilde{t}}, \tilde{t} \in \mathcal{T}, w \in W \tag{5}$$

where $\varepsilon_{I,w}$ and $\varepsilon_{F,w}$ are error terms following certain random distribution. $\varepsilon_{I,w}$ and $\varepsilon_{F,w}$ may follow the same or different random distributions. Function $V(\cdot)$ computes the systematic disutility which is the deterministic part of the perceived disutility. It is modeled to be travel time, based on received information, plus the arrival early/late penalty. The arrival early/late penalty can be modeled in different ways. This paper adopts the one from Han et al. (2019) for the computation in numerical studies. Then, we have the function $V(\cdot)$ defined as follows, where the class of travelers is not differentiated for simplification.

$$V(\phi,t) = \phi + \begin{cases} \mu_1(t+\phi-TA_w)^2 & if\ t+\phi < TA_w \\ \mu_2(t+\phi-TA_w)^2 & if\ t+\phi \geq TA_w \end{cases} \quad w \in W \tag{6}$$

where $\phi$ represents the travel time indicated by either type of information, $TA_w$ represents the target arrival time of OD pair $w$. Penalty parameters $\mu_1$ and $\mu_2$ satisfy $0 < \mu_1 < 1 < \mu_2$, implying that travelers are more sensitive to late arrival. Denote $\mathbf{\Psi}_{I,\tilde{t}}$ and $\mathbf{\Psi}_{F,\tilde{t}}$ as the systematic disutility for later use, i.e.,

$$\Psi_{I,\tilde{t}}(p,t) = V(\phi_{I,\tilde{t}}(p),t), \qquad p \in \mathcal{P}_w, t \in \mathcal{T}_{\tilde{t}}, \tilde{t} \in \mathcal{T}, w \in W \tag{7}$$

$$\Psi_{F,\tilde{t}}(p,t) = V(\phi_{F,\tilde{t}}(p,t),t), \qquad p \in \mathcal{P}_w, t \in \mathcal{T}_{\tilde{t}}, \tilde{t} \in \mathcal{T}, w \in W. \tag{8}$$

Note that, since the instantaneous information provides the travel times for the current time interval $\tilde{t} \in \mathcal{T}$ only, when computing the columns with $t > \tilde{t}$ in $\mathbf{\Psi}_{I,\tilde{t}}$, $\phi_{I,\tilde{t}}(p)$ is used for all $t > \tilde{t}$. Namely, the travelers receiving instantaneous information perceive disutility for all future departure time choices based on the current latest travel time information only. However, the travelers receiving forecast information can perceive disutility for any future departure time choice $t > \tilde{t}$ based on the corresponding forecast travel time associated with $t$ in $\mathbf{\Phi}_{F,\tilde{t}}$. This explains why $\phi_{I,\tilde{t}}(p)$ is used in Eq. (7) and $\phi_{F,\tilde{t}}(p,t)$ is used in Eq. (8).

Given the perceived disutility of all SPDT choices, travelers make tentative choices. This is done by discrete choice models assuming that travelers are minimizing their perceived disutility when making choices. In numerical experiments, we adopt the logit model, i.e., assuming random error terms follow i.i.d. Gumbel distribution. Other discrete choice models are applicable as well.

The logit-based discrete choice model computes the probability of choosing each SPDT choice for travelers. Below presents the well-known model, in the context of this paper, for completeness. It defines the probability of choosing one SPDT choice of path $p \in \mathcal{P}_w$ and departure time $t \in \mathcal{T}_{\tilde{t}}$, at time $\tilde{t} \in \mathcal{T}$, as

$$\mathbb{P}_{p,t} = \frac{e^{-\theta * \Psi(p,t)}}{\sum_{p \in \mathcal{P}_w, t \in \mathcal{T}_{\tilde{t}}} e^{-\theta * \Psi(p,t)}}, \forall p \in \mathcal{P}_w, w \in W, t \in \mathcal{T}_{\tilde{t}} \tag{9}$$

where the dispersion parameter $\theta$ represents the travelers' sensitivity to travel cost (disutility). When $\theta \to 0$, the choice probability assignment is totally random, regardless of travel cost; whilst if $\theta \to \infty$, travelers are minimizing their travel cost.

Based on the Law of Large Numbers, the probabilities computed by the logit model, can be multiplied by the total demand to represent the number of travelers who make the corresponding choices. Therefore, as similarly presented by the matrices *g-l* in the illustrative example, we will use matrices to mathematically represent the tentative choices made by travelers. The matrices, with rows representing paths and columns representing time intervals, contain the values of the number of travelers who make the corresponding SPDT choices. Accordingly, such matrices also represent tentative path departures. Denote matrices $\tilde{\mathbf{h}}_{I,\tilde{t}}$ and $\tilde{\mathbf{h}}_{F,\tilde{t}}$ as the tentative path departures, at $\tilde{t} \in \mathcal{T}$, of the two classes of travelers receiving instantaneous and forecast information, respectively.

Recall in Section 4.1, M is used to denote the mapping from information to tentative path departures. The mapping M is formulated as follows. Again, the class of travelers is not differentiated for simpleness in notations, and the formulation could be used for either class of travelers.

$$M(\mathbf{\Phi}_{\tilde{t}}, \mathbf{d}_{\tilde{t}}) = \tilde{\mathbf{h}}_{\tilde{t}} = \left[\tilde{h}_p(t), p \in \mathcal{P}, t \in \mathcal{T}_{\tilde{t}}\right] \tag{10}$$

where $\tilde{h}_p(t) = \mathbb{p}_{p,t} * d_{\tilde{t},w}, p \in \mathcal{P}_w, w \in W$. The input $\mathbf{\Phi}_{\tilde{t}}$, i.e., the information received at $\tilde{t}$, is used for computation in $\mathbb{p}_{p,t}$ where $\Psi(p,t)$ is obtained based on received information as presented before.

Then, $\tilde{\mathbf{h}}_{I,\tilde{t}}$ and $\tilde{\mathbf{h}}_{F,\tilde{t}}$ can be expressed as follows.

$$\tilde{\mathbf{h}}_{I,\tilde{t}} = M(\mathbf{\Phi}_{I,\tilde{t}}, \mathbf{d}_{I,\tilde{t}}), \mathbf{d}_{I,\tilde{t}} = \left\{ (d_{I,\tilde{t},w}) \middle| d_{I,\tilde{t},w} = d_{I,w} - \sum_{t=1}^{t=\tilde{t}-1} \tilde{\mathbf{h}}_{I,t,p \in \mathcal{P}_w}(t) \right\} \tag{11}$$

$$\tilde{\mathbf{h}}_{F,\tilde{t}} = M(\mathbf{\Phi}_{F,\tilde{t}}, \mathbf{d}_{F,\tilde{t}}), \mathbf{d}_{F,\tilde{t}} = \left\{ (d_{F,\tilde{t},w}) \middle| d_{F,\tilde{t},w} = d_{F,w} - \sum_{t=1}^{t=\tilde{t}-1} \tilde{\mathbf{h}}_{F,t,p \in \mathcal{P}_w}(t) \right\} \tag{12}$$

So far, we have presented the mapping from information to tentative path departures. Next, we present the expression of realized (choices) path departures, selected from the tentative path departures. As mentioned before and in the illustrative example, at time $\tilde{t} \in \mathcal{T}$, only the tentative choices or path departures at the current time will be realized since the tentative choices for future departure times will be updated as new information becomes available as time evolves. Therefore, we have the realized path departures, denoted as $\mathbf{h}_I$ and $\mathbf{h}_F$ for the two classes, expressed as follows.

$$\mathbf{h}_I(\tilde{t}) = \tilde{\mathbf{h}}_{I,\tilde{t}}(\tilde{t}), \qquad \mathbf{h}_F(\tilde{t}) = \tilde{\mathbf{h}}_{F,\tilde{t}}(\tilde{t}), \qquad \tilde{t} \in \mathcal{T} \tag{13}$$

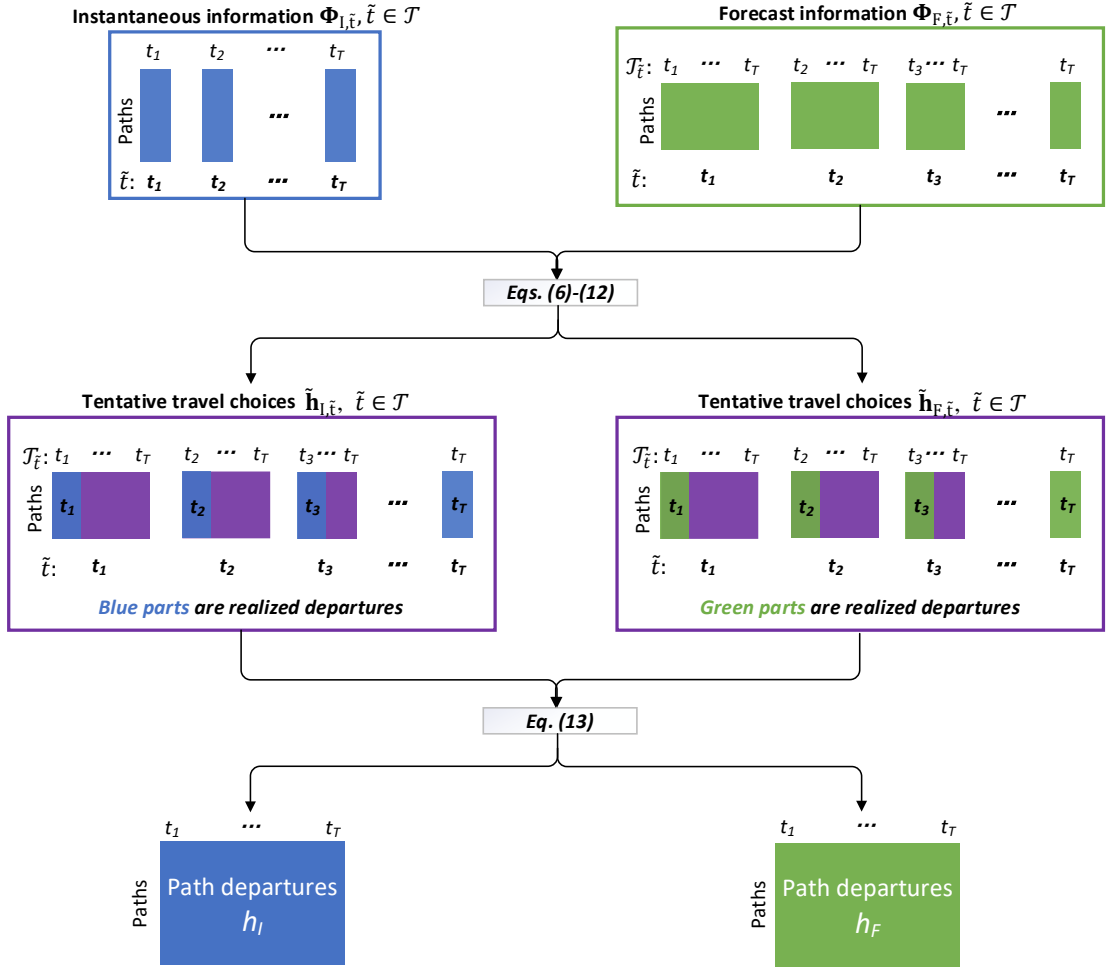

Figure 4. Travel choice (path departures) mapping

Figure 4 shows the workflow of travel choice mapping, which maps from the provided information to realized travel choices or path departures. Travelers make tentative SPDT choices based on the received information, modeled by Eqs. (6)-(12). Such choices are updated as time evolves given the real-time updated travel time information. And only the tentative SPDT choices made at the current time interval will be realized at each time $\tilde{t} \in \mathcal{T}$, as represented by Eq. (13).

*4.3 Equilibrium formulation*

So far, we have presented the proposed mappings in the closed loop of Figure 1. This subsection presents the dynamic traffic equilibrium characterized by the closed loop and formulates it as a fixed-point problem.

It is modeled that travelers aim to minimize the perceived disutility when making tentative travel choices given the provided information at each time interval. Considering stochastic travel behaviors and two travelers classes receiving different real-time updated information, this paper defines the dynamic traffic equilibrium under the proposed information provision mechanism as Dynamic Stochastic User Equilibrium with Dynamic Heterogeneous Information (DSUE-DHI), as follows.

**Definition 1.** The path departures $\mathbf{h}_I^* \in \Lambda_I$ and $\mathbf{h}_F^* \in \Lambda_F$ with $\mathbf{h}^* = \mathbf{h}_I^* + \mathbf{h}_F^*$ are at DSUE-DHI with SPDT choices if travelers minimize their perceived disutility when making travel choices, $\tilde{\mathbf{h}}_{I,\tilde{t}}^*(\tilde{t})$ and $\tilde{\mathbf{h}}_{F,\tilde{t}}^*(\tilde{t})$, at $\tilde{t} \in \mathcal{T}$. In other words, at equilibrium, no travelers can further reduce their perceived disutility by unilaterally changing their SPDT choice at each time interval, given the received information.

The DSUE-DHI can be formulated as the solution $(\mathbf{h}_I^*, \mathbf{h}_F^*)$ of the following fixed-point problem.

Find $\mathbf{h}_I^* \in \Lambda_I, \mathbf{h}_F^* \in \Lambda_F$ that satisfy

$$\begin{cases} \mathbf{h}_I^*(\tilde{t}) = \mathrm{M}\big(\boldsymbol{\Phi}_{I,\tilde{t}}(\mathbf{h}^*), \mathbf{d}_{I,\tilde{t}}\big)(\tilde{t}), \tilde{t} \in \mathcal{T} \\ \mathbf{h}_F^*(\tilde{t}) = \mathrm{M}\left(\boldsymbol{\Phi}_{F,\tilde{t}}\left(\boldsymbol{\hbar}_{\tilde{t}}\left(\boldsymbol{\Phi}_{I,\tilde{t}}(\mathbf{h}^*)\right)\right), \mathbf{d}_{F,\tilde{t}}\right)(\tilde{t}), \tilde{t} \in \mathcal{T} \\ \mathbf{h}^* = \mathbf{h}_I^* + \mathbf{h}_F^* \end{cases} \quad (14)$$

Eq. (14) can be further written as the following with path departures $\mathbf{h}^*$ without $\mathbf{h}_I^*$ and $\mathbf{h}_F^*$:

$$\mathbf{h}^*(\tilde{t}) = \mathrm{M}\big(\boldsymbol{\Phi}_{I,\tilde{t}}(\mathbf{h}^*), \mathbf{d}_{I,\tilde{t}}\big)(\tilde{t}) + \mathrm{M}\left(\boldsymbol{\Phi}_{F,\tilde{t}}\left(\boldsymbol{\hbar}_{\tilde{t}}\left(\boldsymbol{\Phi}_{I,\tilde{t}}(\mathbf{h}^*)\right)\right), \mathbf{d}_{F,\tilde{t}}\right)(\tilde{t}), \tilde{t} \in \mathcal{T} \quad (15)$$

From Eq. (15), the path departures matrix $\mathbf{h}^*$ undergoes a series of mappings, ultimately converging to itself, representing a fixed-point problem. Figure 5 shows the fixed-point relationship, with the closed loop composed of the three mappings. In particular, the information generation mapping involves the mapping M (Eq. (1)) for generating forecast path departures and the DNL mapping for generating the forecast information. Therefore, the complexity arises not only from the dynamic updates of travel time information, travel choices, and traffic condition, but also from the embedded mappings in the information generation.

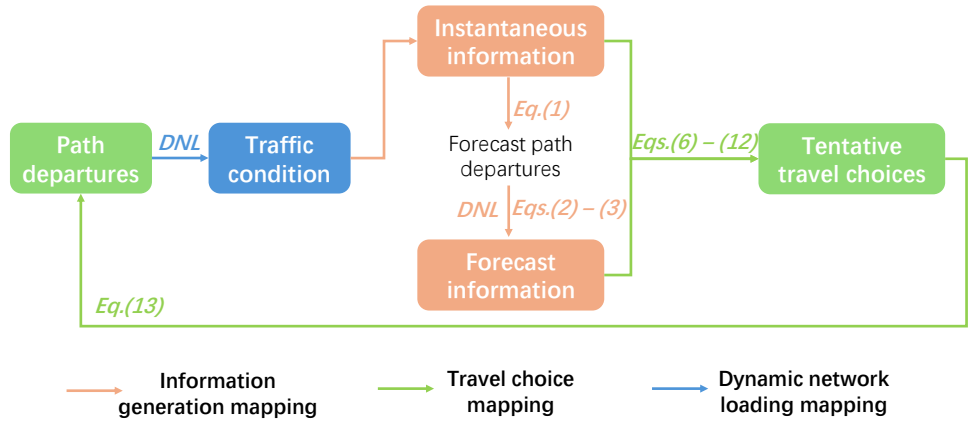

Figure 5. The fixed-point problem

## 5. Properties

This section proposes several properties of the proposed DSUE-DHI, including the equivalence between the fixed-point problem and DSUE-DHI, solution existence, its relationship with the classical single-class dynamic stochastic user equilibrium (DSUE), and the analysis of information accuracy.

This paper stipulates the provided travel time information as *accurate* when it is identical to the realized travel time. *Information accuracy* is measured by the difference between the provided travel time information and the realized travel time. A smaller difference indicates higher information accuracy, and a larger difference implies lower accuracy.

**Proposition 1**. Path departures of the two groups of travelers $\mathbf{h}_I \in \Lambda_I$ and $\mathbf{h}_F \in \Lambda_F$ with $\mathbf{h} = \mathbf{h}_I + \mathbf{h}_F$ are at DSUE-DHI if and only if $\mathbf{h}_I$ and $\mathbf{h}_F$ is a solution to the fixed-point problem (14).

**Proof.** *Necessity:* If path departures $\mathbf{h}_I \in \Lambda_I$ and $\mathbf{h}_F \in \Lambda_F$ are at DSUE-DHI, according to Definition 1, travelers seek to minimize the perceived disutility when making choices. At equilibrium, no one can further reduce her/his perceived disutility by unilaterally changing her/his SPDT choices at each time interval. Based on the random utility theory and the definition of discrete choice model, the path departures $\mathbf{h}_I$ and $\mathbf{h}_F$ resulted from the provided information must be computed as what is defined in the travel choice mapping. If

$$\mathbf{h}_I(t) \neq M\big(\mathbf{\Phi}_{I,t}(\mathbf{h}), \mathbf{d}_{I,t}\big)(t), \exists t \in \mathcal{T}$$

or

$$\mathbf{h}_F(t) \neq M\big(\mathbf{\Phi}_{F,t}(\mathbf{h}), \mathbf{d}_{F,t}\big)(t), \exists t \in \mathcal{T}$$

holds, then $\mathbf{h}_I$ or $\mathbf{h}_F$ are not the path departures solved by the travel choice mapping given the provided travel time information. Namely, there are travelers whose perceived disutility is not minimized when making choices, based on the random utility theory and the definition of discrete choice model. They can change their choices to reduce the perceived disutility with the provided information. This contradicts Definition 1. Therefore, Eq. (14) must hold at DSUE-DHI. That is, $\mathbf{h}_I$ and $\mathbf{h}_F$ must be a solution to the proposed fixed-point problem.

*Sufficiency:* If $\mathbf{h}_I$ and $\mathbf{h}_F$ is a solution to fixed-point problem (14), it is implied that, given the provided information resulted from $\mathbf{h}_I$ and $\mathbf{h}_F$, the path departures solved by the proposed travel choice mapping must be equal to $\mathbf{h}_I$ and $\mathbf{h}_F$ themselves. In the travel choice mapping, it is defined that travelers are minimizing their perceived disutility when making choices. Given this, the fixed-point relationship in (14) implies that, with $\mathbf{h}_I$ and $\mathbf{h}_F$, no one can further reduce her/his perceived disutility by unilaterally changing her/his SPDT choices at each time interval, which is at DSUE-DHI. ∎

**Proposition 2**. The solution to the fixed-point problem exists if the applied DNL model and the discrete choice model are continuous.

**Proof.** According to Brouwer's fixed-point theorem (Park, 1999), for any continuous function $f$ mapping a compact convex set to itself, there is a point $x^*$ such that $f(x^*) = x^*$. For fixed-point problem (15), $\mathbf{h} = \mathbf{h}_I + \mathbf{h}_F, \mathbf{h}_I \in \Lambda_I, \mathbf{h}_F \in \Lambda_F$. Sets $\Lambda_I$ and $\Lambda_F$ are closed, bounded, and convex. Hence, the feasible set of $\mathbf{h}$ is a compact convex set. The continuity of the underlying function $f$, which is the composite mapping involving the information generation mapping, travel choice mapping, and the DNL mapping, is proved as follows.

Since mapping M is continuous based on the continuity of the disutility function and the discrete choice model, the travel choice mapping is continuous. Given the continuity of the DNL model, we only need to prove the continuity of the information generation mapping for proving this Proposition.

Denote $\bar{t}_p(\tilde{t})$ as the travel time on path $p$ when departure time is $\tilde{t}$. Let matrix $\mathbf{Z}^{p,\tilde{t}} = (\zeta_{a,t}, a \in L, t \in \mathcal{T})$ where $\zeta_{a,t} = 1$ if path $p$ uses link $a$ and flows departing on path $p$ at time $\tilde{t}$ arrive link $a$ at time $t$; and 0 otherwise. The DNL model solves the travel times on each link for all departure time intervals, denoted as $\overline{\mathbf{T}} = (\bar{T}_{a,t}, a \in L, t \in \mathcal{T})$. The path travel time $\bar{t}_p(\tilde{t})$ is the Frobenius inner product of $\mathbf{Z}^{p,\tilde{t}}$ and $\overline{\mathbf{T}}$, i.e., $\bar{t}_p(\tilde{t}) = \langle \mathbf{Z}^{p,\tilde{t}}, \overline{\mathbf{T}} \rangle_\mathrm{F}$. As the path travel time obtained from the DNL model is assumed to be continuous with respect to path departures, one can conclude that the link travel time $\overline{\mathbf{T}}$ is continuous.

The instantaneous travel time information, $\mathbf{\Phi}_{\mathrm{I},\tilde{t}} = \Delta^T * \overline{\mathbf{T}}(\tilde{t})$, is continuous due to the continuity of $\overline{\mathbf{T}}$. According to Eqs. (2)-(3), the forecast travel time information $\mathbf{\Phi}_{\mathrm{F},\tilde{t}}$ is directly obtained from the DNL model with the input $\hbar_{\tilde{t}}$ that is continuous with respect to $\mathbf{h}$. Here, $\hbar_{\tilde{t}}$ is continuous with respect to $\mathbf{h}$, since mapping $\mathbb{M}$ is continuous. Therefore, given the continuity of the DNL model, the forecast travel time information $\mathbf{\Phi}_{\mathrm{F},\tilde{t}}$ is also continuous. The continuity of the information generation mapping has been proved.

Therefore, the underlying function $f$, a composite mapping of continuous mappings, is also continuous. By Brouwer's fixed-point theorem, there exists at least one solution to the fixed-point problem. Note that Brouwer's fixed-point theorem is applicable even though the adopted discrete choice model never yields a $\mathbf{h}$ with zero path flows, namely, the solution set $\mathbf{h}^*$ is not closed. From engineering point of view, there must exist a threshold $\delta > 0$ that is sufficiently small such that $\mathbf{h} \geq \delta$ always holds for all path flows when the solution achieves a predefined precision. As a result, the set of $\mathbf{h}$ defined upon $\mathbf{h} \geq \delta$ is a compact convex set. ∎

**Remark 1**. The uniqueness of DSUE-DHI cannot be guaranteed due to the complexities arising from two main aspects: (1) the complex flow propagation dynamics along the space-and-time dimensions in DNL, and (2) the multi-class nature considered in the proposed model. Recognizing that the DNL model lacks a closed-form formula, its mathematical properties remain largely unknown (Han et al., 2019). Long et al. (2015) postulate that DSUE with SPDT choices may not have a unique solution due to the non-monotonicity of the travel delay function. Moreover, even in static traffic equilibrium, studies such as Nagurney (2000) and Yang and Huang (2004) showed that a multi-class traffic system could have multiple equilibria. The non-uniqueness of multi-class dynamic traffic equilibrium has also been noted by Huang and Lam (2003) in their proposed model. Therefore, the uniqueness of DSUE-DHI, considering multi-class information provision with complex dynamic structures, cannot be guaranteed.

Consider the classical single-class dynamic stochastic user equilibrium without any specific information provision involving the same DNL model, disutility function, and discrete choice model as used in DSUE-DHI. Label this single-class counterpart as DSUE in the rest of the paper for simplicity. In addition, in this paper, a network is referred to as uncongested when it experiences free flow travel time.

**Proposition 3**. DSUE is a special case of the proposed DSUE-DHI. DSUE-DHI reduces to be DSUE when the network is uncongested.

**Proof.** When the network is uncongested, the instantaneous travel time information and forecast travel time information for all future departure times are identical and accurate as the realized travel time which is free-flow travel time. There is essentially one class of travelers in the system and the information provided at each time does not change. Travelers make choices based on their perceived disutility which is calculated upon the provided realized travel time information. Essentially, $\boldsymbol{\Phi}_{I,\tilde{t}}(\mathbf{h}^*)$ and $\boldsymbol{\Phi}_{F,\tilde{t}}(\mathbf{h}^*)$ are identical for all $\tilde{t} \in \mathcal{T}$, since there is no congestion evolves, and travelers' choices do not change over time. And the proposed Definition 1 reduces to be the following fixed-point problem.

$$\mathbf{h}^* = \mathrm{M}(\boldsymbol{\Phi}(\mathbf{h}^*), \mathbf{d})$$

That is, at equilibrium, no one can further reduce her/his perceived disutility by unilaterally changing her/his SPDT choice, given the realized travel time information. This is exactly equivalent to the classical DSUE where the random term in disutility is directly added upon the realized travel cost. ∎

**Remark 2**. Based on Proposition 3, the proposed DSUE-DHI distinguishes itself from DSUE in congested scenarios but reduces to DSUE under the uncongested condition. In addition, DSUE-DHI aligns with the reality that the travel time information tends to be accurate in uncongested scenarios but could be inaccurate and play a more significant role in affecting network performance in congested scenarios.

**Proposition 4**. If the mapping $DNL$ is monotone, the accuracy of forecast information decreases as more travelers receive it. In particular, at DSUE-DHI, when no travelers receive forecast information, i.e., the system contains only the class receiving instantaneous information, the strategic forecast information for departure time $\tilde{t} \in \mathcal{T}$ provided at $\tilde{t} \in \mathcal{T}$ is identical to the realized travel time. Namely, in this scenario, the latest forecast information provided at each time interval is accurate for departures at that time.

**Proof.** At time $\tilde{t} \in \mathcal{T}$, at DSUE-DHI, the forecast travel time for departure time $\tilde{t}$ is

$$\boldsymbol{\Phi}_{F,\tilde{t}}(\tilde{t}) = [\phi_{F,\tilde{t}}(p,\tilde{t}), p \in \mathcal{P}] \text{ with } \phi_{F,\tilde{t}}(p,\tilde{t}) = DNL(\boldsymbol{\hslash}_{\tilde{t}}^*)(p,\tilde{t})$$

$$\boldsymbol{\hslash}_{\tilde{t}}^*(t) = \begin{cases} \mathbf{h}^*(t), & if\ t < \tilde{t} \\ \bar{\mathbf{h}}_{\tilde{t}}^*(t), & if\ t \geq \tilde{t} \end{cases}$$

$$\bar{\mathbf{h}}_{\tilde{t}}^*(t) = \mathrm{M}(\boldsymbol{\Phi}_{I,\tilde{t}}(\mathbf{h}^*), \mathbf{d}_{\tilde{t}})(t)$$

And the realized travel time for departure time $\tilde{t}$ is

$$DNL(\mathbf{h}^*)(\tilde{t}).$$

Let $\lambda \in [0,1]$ represent the proportion of the travelers receiving instantaneous information, then

$$\mathbf{h}^*(\tilde{t}) = \lambda \mathrm{M}(\boldsymbol{\Phi}_{I,\tilde{t}}(\mathbf{h}^*), \mathbf{d}_{\tilde{t}})(\tilde{t}) + (1-\lambda)\mathrm{M}(\boldsymbol{\Phi}_{F,\tilde{t}}(\mathbf{h}^*), \mathbf{d}_{\tilde{t}})(\tilde{t}).$$

When there is no traveler receiving forecast information, i.e., $\lambda = 1$, we have $\mathbf{h}^*(\tilde{t}) = M(\mathbf{\Phi}_{I,\tilde{t}}(\mathbf{h}^*), \mathbf{d}_{\tilde{t}})(\tilde{t}) = \bar{\mathbf{h}}_{\tilde{t}}^*(\tilde{t})$. Therefore, $\boldsymbol{h}_{\tilde{t}}^*(t) = \mathbf{h}^*(t), t \leq \tilde{t}$, holds. Because the travel time for departure time $\tilde{t}$ depends on the departure dynamics in $t \leq \tilde{t}$ only and has nothing to do with future departure dynamics, we have $DNL(\boldsymbol{h}_{\tilde{t}}^*)(\tilde{t}) = DNL(\mathbf{h}^*)(\tilde{t})$, implying that the forecast information provided at each time interval is accurate for departures at that time.

If the $DNL$ mapping is monotone, the difference between the forecast travel time $DNL(\boldsymbol{h}_{\tilde{t}}^*)(\tilde{t})$ and realized travel time $DNL(\mathbf{h}^*)(\tilde{t})$ is in positive correlation to the difference between $\bar{\mathbf{h}}_{\tilde{t}}^*(\tilde{t})$ and $\mathbf{h}^*(\tilde{t})$. The relationship between $\bar{\mathbf{h}}_{\tilde{t}}^*(\tilde{t})$ and $\mathbf{h}^*(\tilde{t})$ can be written as Eq.(16). Based on Eq.(16), one can conclude that as more travelers receive forecast information, i.e., as $\lambda$ decreases, the difference between $\bar{\mathbf{h}}_{\tilde{t}}^*(\tilde{t})$ and $\mathbf{h}^*(\tilde{t})$ becomes larger. Hence, the difference between $DNL(\boldsymbol{h}_{\tilde{t}}^*)(\tilde{t})$ and $DNL(\mathbf{h}^*)(\tilde{t})$ becomes larger. That is, the forecast information becomes less accurate as more travelers receive it. The proof is complete.

$$\mathbf{h}^*(\tilde{t}) = \lambda \bar{\mathbf{h}}_{\tilde{t}}^*(\tilde{t}) + (1-\lambda) M(\mathbf{\Phi}_{F,\tilde{t}}(\mathbf{h}^*), \mathbf{d}_{\tilde{t}})(\tilde{t}) \tag{16}$$

**Proposition 5.** At equilibrium of DSUE-DHI, the accuracy of forecast information is higher than or equal to that of instantaneous information in the Euclidean norm, if the $DNL$ mapping is continuous and convex. Namely, the following inequality holds if the $DNL$ mapping is continuous and convex:

$$\|\mathbf{\Phi}_I(\mathbf{h}^*) - \bar{\mathbf{t}}(\mathbf{h}^*)\| \geq \|\mathbf{\Phi}_F(\mathbf{h}^*) - \bar{\mathbf{t}}(\mathbf{h}^*)\| \tag{17}$$

where $\bar{\mathbf{t}}$ contains the realized travel times for all pairs of departure time $t \in \mathcal{T}$ and path $p \in \mathcal{P}$. $\mathbf{\Phi}_I$ contains the instantaneous travel time information provided at each time $t \in \mathcal{T}$ for each path $p \in \mathcal{P}$, and $\mathbf{\Phi}_F$ contains the forecast travel time information provided at each time $t \in \mathcal{T}$ for departure time $t$ and each path $p \in \mathcal{P}$. $\mathbf{h}^*$ represents the path departures at DSUE-DHI.

**Proof.** See Appendix B.

Note that the properties presented in this section are independent of the specific DNL model and discrete choice model used. Other DNL and discrete choice models can be applied. In the numerical experiments, the logit-based discrete choice model and the DNL model proposed in Han et al. (2019) are adopted. The logit-based discrete choice model is known to be continuous, and the continuity of the DNL model has been proved by Han et al. (2016). However, the monotonicity and convexity of the adopted DNL model remain unclear and challenging to prove due to its non-closed form, as discussed in Han et al. (2019). Nevertheless, the results in Section 7 show numerical consistency with the propositions presented in this section.

## 6. Solution Algorithm

The Method of Successive Averages (MSA) has been extensively used to solve fixed point problems. MSA is an iterative method that uses the weighted average of the solution in the previous step and its derived function value, i.e., the auxiliary point, as the new solution in each iteration. The convergence of the MSA requires two conditions for step size determination: (1) the step size converges to zero, and (2) the summation of all step sizes converges to infinity (Robbins and Monro, 1951, Blum, 1954). Note that the convergence speed of MSA is slow due to two reasons, as pointed out in the literature (Liu et al., 2009, Szeto et al., 2011). First, the predetermined step size is too large at some iterations such that the updated solution is farther away from the optimum. Second, the predetermined step size is too small when the current solution is close to the optimum. Liu et al. (2009) tackles the slow convergence of MSA by developing a self-regulated averaging method (SRAM), in which the step sizes vary depending on the distance between the intermediate solution and auxiliary point. This paper adopts the SRAM to solve the proposed fixed-point problem.

Denote $y^k$ as the auxiliary point with the temporary solution $\mathbf{h}^k$ at the k-*th* iteration. The solution algorithm applies the SRAM presented as follows.

**Step 1. (Initialization)** Set iteration number $k = 1$. Initialize step size $\beta^1 = 1$ and path departures $\mathbf{h}^1$. Set the maximum number of iterations $k_{max}$, the convergence tolerance $\xi$, and the step size parameters $\Gamma$ and $\gamma$.

**Step 2. (Update the mapping)** Calculate $y^k$ using Eq. (15) with the input $\mathbf{h}^k$.

*Step 2.1.* Obtain the instantaneous travel time information $\mathbf{\Phi}_{I,t}(\mathbf{h}^k)$ and the forecast travel time information $\mathbf{\Phi}_{F,t}(\mathbf{h}^k)$, using the DNL procedure and the information generation mapping: Eqs. (1)-(3), for $t \in \mathcal{T}$.

*Step 2.2.* Obtain the resulting path departures $\mathbf{h}_I$ and $\mathbf{h}_F$ using the travel choice mapping: Eqs. (6)-(13), for $t \in \mathcal{T}$. Let $y^k = \mathbf{h}_I + \mathbf{h}_F$.

**Step 3. (Check convergence)** Stop if (i) $k > k_{max}$, or (ii) $\|\mathbf{h}^k - y^k\|^2 / \|\mathbf{h}^k\|^2 \leq \xi$.

**Step 4. (Determine step size)** The self-regulated step size $\alpha^k = 1/\beta^k$, and $\beta^k$ is updated by:

$$\beta^k = \begin{cases} \beta^{k-1} + \Gamma, \Gamma > 1, & if \ \|\mathbf{h}^k - y^k\| \geq \|\mathbf{h}^{k-1} - y^{k-1}\| \\ \beta^{k-1} + \gamma, 0 < \gamma < 1, & if \ \|\mathbf{h}^k - y^k\| < \|\mathbf{h}^{k-1} - y^{k-1}\| \end{cases}$$

**Step 5. (Update the path departures)** $\mathbf{h}^{k+1} = \mathbf{h}^k + \alpha^k \cdot (y^k - \mathbf{h}^k)$. $k = k + 1$ and go to Step 2.

## 7. Numerical Experiment

This section demonstrates the proposed model and investigates the influences of proposed travel time information provision using the Anaheim network, which contains 416 nodes, 914 links, and 1406 OD pairs. The network is shown in Figure 6. The network structure and link characteristics are directly adopted from Han et al. (2019). Given that the proposed DSUE-DHI formulation is path-based, path enumeration is applied. In particular, paths are enumerated with constraints on free-flow travel time and path length to avoid detours and unreasonable paths, using the method presented by Van der Zijpp and Catalano (2005). In this example, the time horizon of the analysis is five hours, with each discrete time interval being two minutes. The parameters $\mu_1$ and $\mu_2$ in Eq. (6) are adopted from Han et al. (2019) as 0.8 and 1.2, respectively. The dispersion parameter in the logit-based discrete choice model $\theta$ ranges from 0.1 to 2, representing different levels of travelers' sensitivity to travel cost (disutility). The proportion of travelers who receive the forecast information ranges from 0% to 100%. For the solution algorithm, the threshold value $\xi$ of the stopping condition is set to be $10^{-4}$; the step size parameters are set to be $\Gamma=1.1$ and $\gamma=0.2$; and the maximum number of iterations is 100.

This research defines *Regular Class* travelers as those who receive instantaneous traffic information and *Strategic Class* travelers as those who receive strategically forecast traffic information. In the rest of the paper, these two classes of travelers are labeled as '*regular travelers*' and '*strategic travelers*' for simplicity. In addition, we will use the term *Relative Difference* as a measurement between two scalars or vectors in the presentation of the numerical study. The *Relative Difference* between $A$ and $B$ is defined as $\frac{A-B}{B}$ if they are scalars and $\frac{\|A-B\|}{\|B\|}$ if they are vectors.

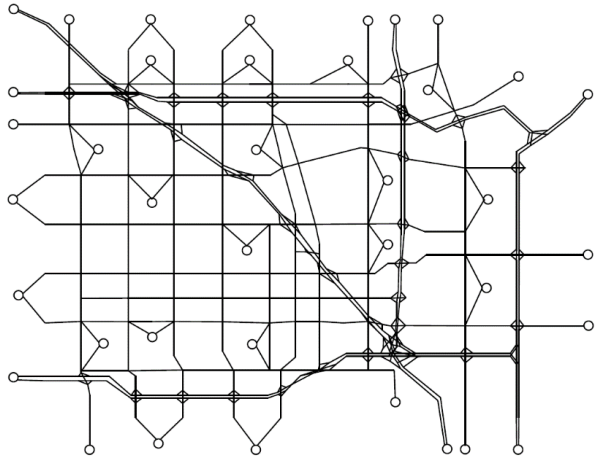

Figure 6. Anaheim Network

The rest of this section will first present a base case in Section 7.1, followed by the evaluation of the effects of the dispersion parameter and the proportion of strategic travelers on the equilibrium in Section 7.2. Section 7.3 focuses on the effect of congestion level by presenting a scenario with higher demand. Section 7.4 demonstrates the computational convergence for solving the proposed equilibrium.

*7.1 Base case*

The base case sets the dispersion parameter $\theta=1$ and the proportion of *strategic* travelers as 50%. Figure 7 shows the path departure rates of the two classes of travelers for one OD pair at the base-case DSUE-DHI. In Figure 7, majority of travelers take two paths, and their departure times concentrate between the second and the fourth hour. Comparing the two classes of travelers, one can see that strategic travelers depart earlier, and their departure times spread in a broader time interval than regular travelers. This is because the forecast information can predict future congestion to some extent. Some strategic travelers choose to depart earlier. In essence, the forecast information can help spread traffic to peak shoulder hours in this scenario.

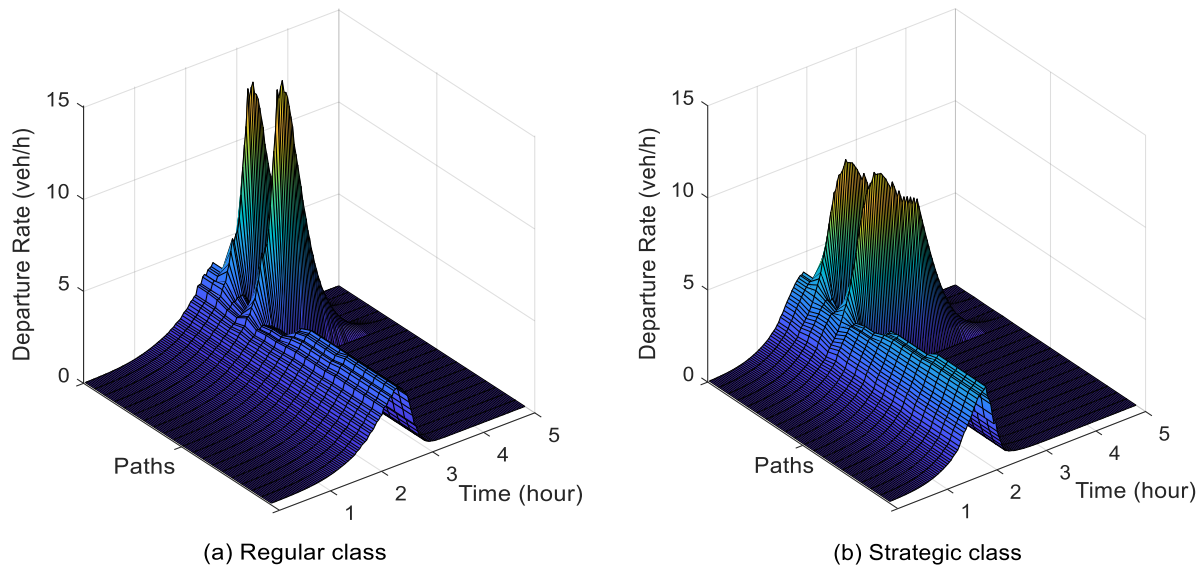

(a) Regular class          (b) Strategic class

Figure 7. Path departure rates for the two classes of travelers for one selected OD pair at DSUE-DHI

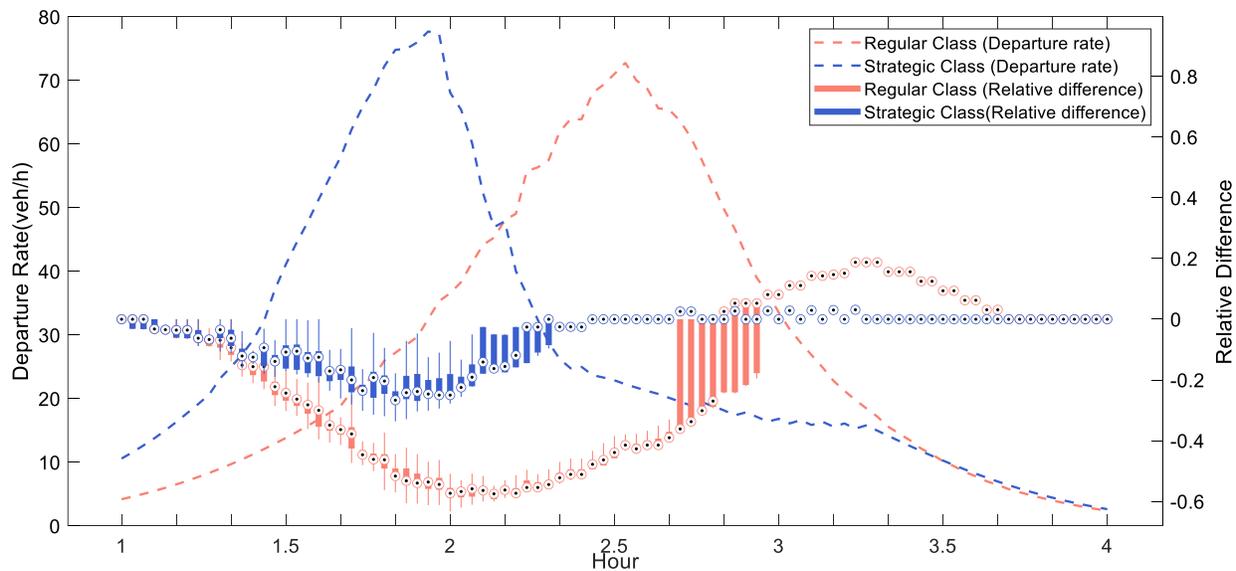

Figure 8. Distribution variation of relative difference between travel time information and realized travel time for the two classes over time (Dashed lines: path departure rates)

Figure 8 shows the distribution variation of the relative difference between travel time information and realized travel time for the two classes in the selected OD pair over time. The departure time intervals displayed range from the end of the first hour to the end of the fourth hour, with the first hour and the last hour in the whole time horizon removed as the warm-up and cool-down period, respectively. The circles with a dot inside represent the median of relative differences among all the travelers who departed at the corresponding departure time. The bottom and top edges of each box indicate the 25th and 75th percentiles, respectively. The line extended from the box ends up at the maximum and down at the minimum, respectively. Dashed lines show the path departure rates over time.

From Figure 8, the forecast information is more accurate and has smaller variations in the relative differences in general. The instantaneous information underestimates travel time by up to 60% and overestimates by up to 20%. Such bounds for the strategic forecast information are significantly smaller, with 30% and 3%, respectively. At the starting and the ending periods, where the network is not congested since most travelers have not departed or have arrived at their destination, the information accuracy is quite high as the relative differences distribute around zero for both classes. In line with the discussion in Remark 2, the instantaneous information and forecast information are identical to the realized travel time at these uncongested departure time intervals. As the congestion increases, the difference between instantaneous information and strategic information in terms of information accuracy shows up and becomes salient, shown as the larger gap from the 1.5 Hour to around the 2.8 Hour.

The provided information tends to underestimate the travel time until the 2.3 Hour and the 3rd Hour, for the regular and strategic class respectively. After the 2.3 Hour, the strategic information is almost accurate. This is because at these time intervals, there are few strategic travelers in the system, so the forecast information accuracy is high according to Proposition 4. After the 3rd Hour, the instantaneous information overestimates the travel time for a certain period. This is because, travelers gradually arrive at the destination, and the congestion ebbs, while the instantaneous travel time reflects the real-time traffic condition only and cannot predict the decreasing travel time on the subsequent links, resulting in overestimation of realized travel time for trips departed at those time intervals.

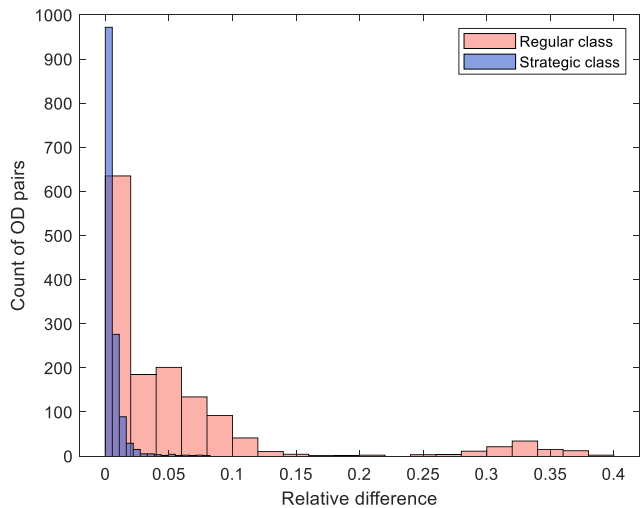

Figure 9. Histograms of the relative difference between informed travel time (*ITT*) and realized travel time (*RTT*) for the two classes among all OD pairs. (*ITT*(*or RTT*) is a vector containing the informed (or realized) travel

times for all SPDT choices)

Figure 9 summarizes the information accuracy distributions for the two classes for all OD pairs, which is quantified by the relative difference between the informed travel time and the realized travel time. Thus, a smaller relative difference means a higher information accuracy or reliability. Figure 9 highlights that the forecast information is much more accurate than instantaneous information, as more OD pairs experience lower relative differences for the strategic class. Such relative differences for strategic travelers are within the 5% range, with the majority close to zero. However, for regular travelers, the relative differences are much larger, up to 40% at the maximum.

Figure 10 shows the comparison between the two classes of travelers in terms of average experienced systematic disutility for each OD pair. We can find that there exist OD pairs in which strategic travelers experience a little higher systematic disutility than regular travelers on average, although they are more likely to gain more reliable travel times from the forecast information. Note that, although the regular travelers of about half of OD pairs gain more benefits on average, their received benefit is limited below 10%. On the other hand, the negative-side long tail of the distribution shown in the figure means that, for the OD pairs where strategic travelers gain more benefits, regular travelers might have to suffer much more in terms of average disutility than strategic travelers (around twice at the maximum).

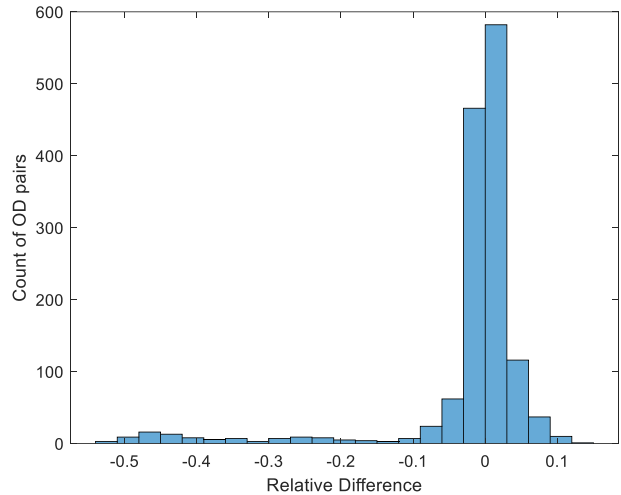

Figure 10. Relative difference between average systematic disutility of strategic class and regular class ((strategic-regular) / regular) – histogram on all OD pairs.

Next, we compare the DSUE (with $\theta = 1$) with the base case of DSUE-DHI, at the system level and individual level. Figure 11 shows the relative difference between DSUE-DHI and DSUE in terms of (a) total systematic disutility and (b) total travel time for all OD pairs. From Figure 11 (a), about half of the OD pairs experience more disutility under DSUE-DHI, while another half of the OD pairs experience less disutility under DSUE-DHI. The relative differences concentrate around zero, meaning there is no significant difference for most OD pairs. Figure 11 (b) shows similar comparison results in terms of total travel time.

Figure 12 compares the distributions of travelers' experienced disutility at DSUE and DSUE-DHI. As shown, the two distributions are similar, while DSUE-DHI performs slightly better than DSUE. Note that this comparison is only for the base case, and the results could vary depending on the parameters' values. Section 7.2 will present the results with different values of parameters.

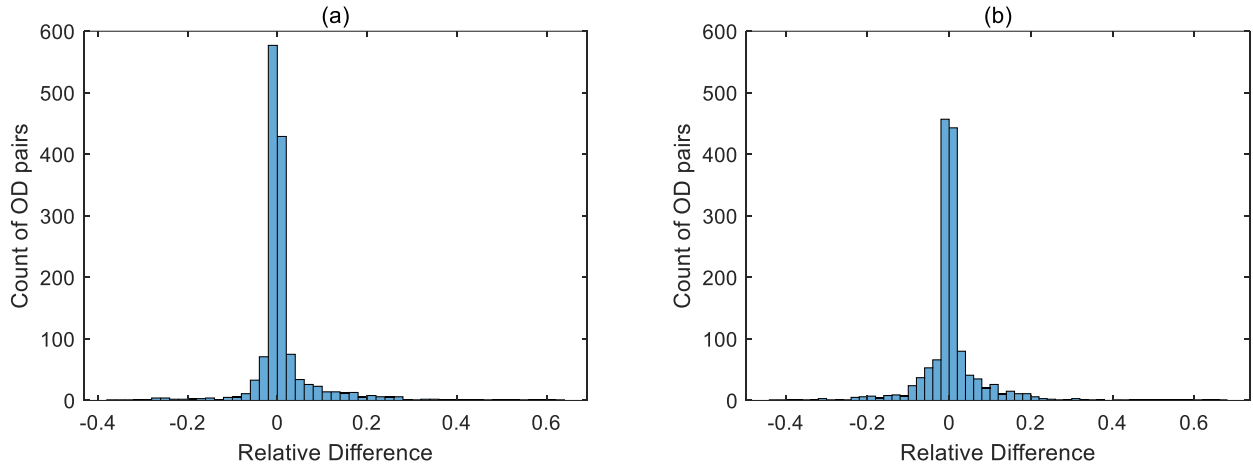

Figure 11. Relative difference between the, (a) total systematic disutility; (b) total travel time, of DSUE-DHI and DSUE for each OD pair – histogram by OD pairs.

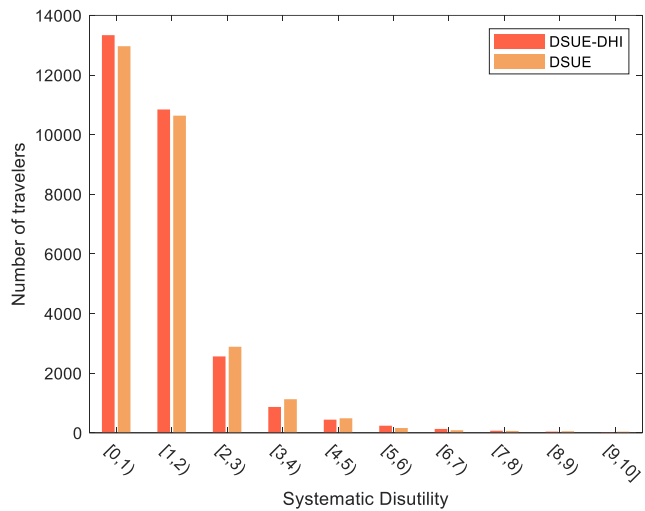

Figure 12. Distributions of travelers' experienced disutility under DSUE and DSUE-DHI

*7.2 Sensitivity analysis*

In this section, we examine the effects of two parameters, dispersion parameter $\theta$ and the proportion of strategic travelers. For the sensitivity analysis of $\theta$, the proportion of strategic travelers keeps as 50%; and for the analysis of the proportion of strategic travelers, the value of $\theta$ keeps as 1.

Figure 13 shows the path departures at DSUE-DHI on a selected path in the same OD pair presented in Figure 7 for four cases where $\theta$ is set to be 0.5, 1, 1.5, and 2, respectively. We can see that as $\theta$ increases, the path departures become more concentrated because travelers are getting more sensitive to travel cost.

Figure 14 presents the relative difference between the informed travel time and realized travel time as $\theta$ changes. We can find that strategic travelers gain more benefits than regular travelers in terms of information accuracy as $\theta$ increases. As $\theta$ increases, the information accuracy decreases for both classes of travelers, while such decrease is much severer for the regular class.

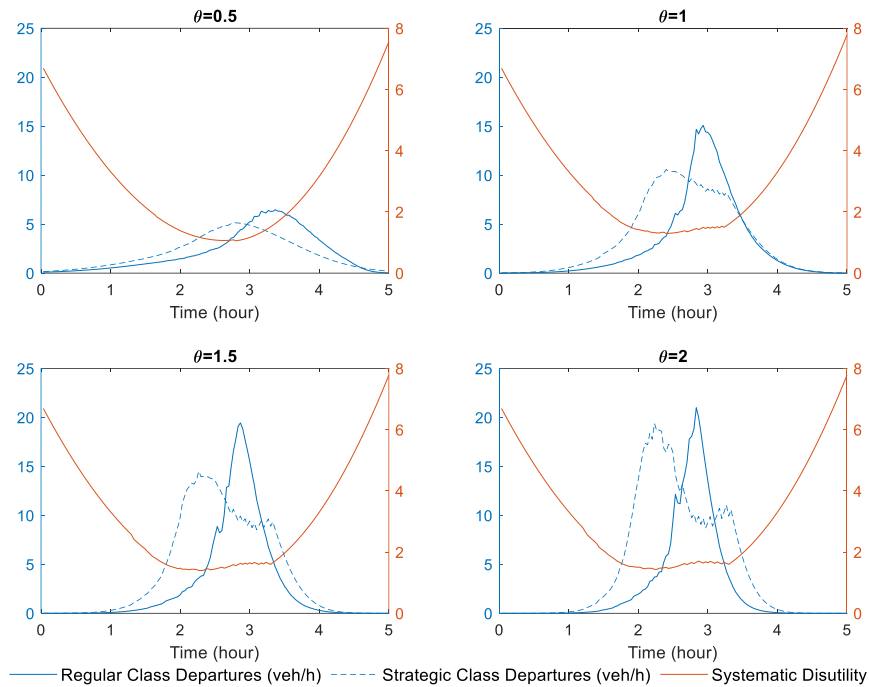

Figure 13. Path departures on a selected path at DSUE-DHI with different values of $\theta$

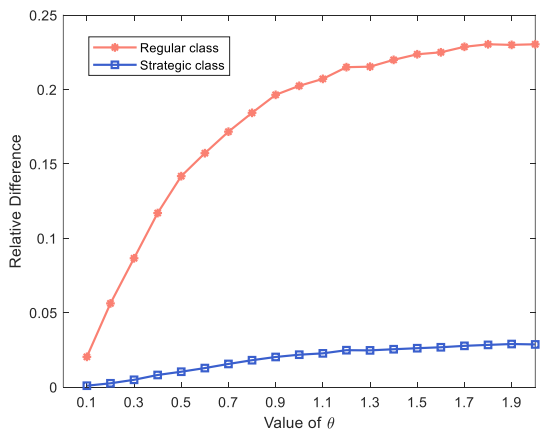

Figure 14. Relative difference between *ITT* and *RTT* as $\theta$ changes

Figure 15 (a) shows the change of average systematic disutility as the value of $\theta$ changes for DSUE-DHI, with DSUE as a reference. We can see that, as $\theta$ increases, the average systematic disutility decreases. Such decrease is larger for the strategic class than the regular class. This implies that, as the travelers' sensitivity to travel cost increases, strategic travelers receive more benefits than regular travelers in terms of travel cost. In particular, when compared to the DSUE case, the strategic travelers are better off while the regular travelers are worse off, leading to the whole group (where each class takes a half in this case) a little worse off since the gap with the regular class outweighs that with strategic class. Moreover, the difference between DSUE-DHI and DSUE in terms of average systematic disutility becomes more apparent as travelers are more sensitive to travel costs (as $\theta$ increases).

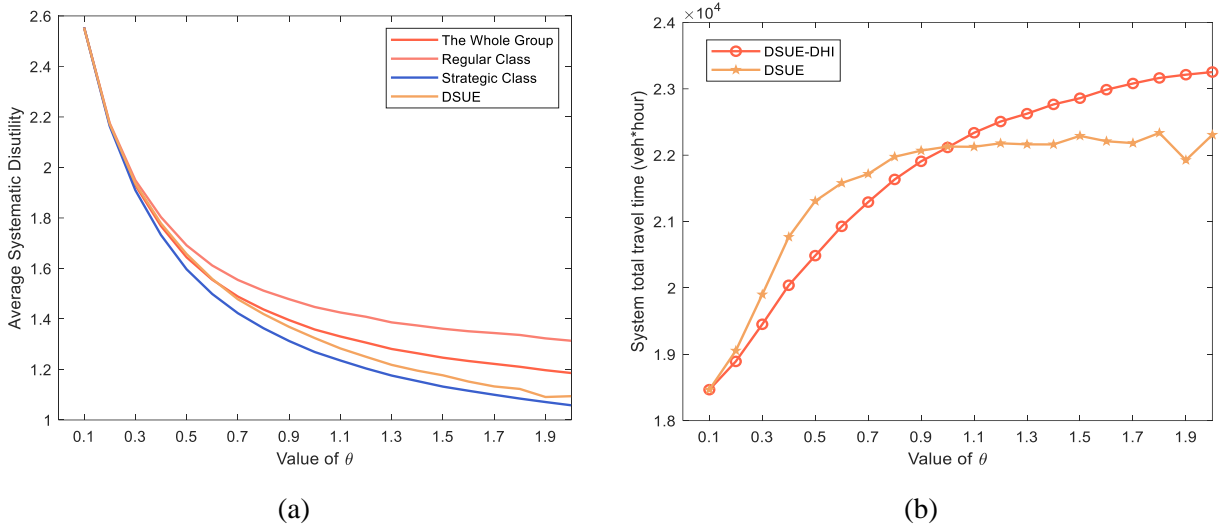

(a) (b)

Figure 15. Changes of, (a) average systematic disutility, (b) system total travel time, as $\theta$ changes

Figure 15 (b) shows the change in network performance in terms of total system travel time as the value of $\theta$ changes for DSUE-DHI, with the DSUE as a reference. As $\theta$ increases, the system total travel time increases in general for both cases. There is a threshold value of $\theta$, around 1, smaller than which the DSUE-DHI performs better; otherwise, the DSUE performs better. This implies that, when travelers' sensitivity to travel cost is small, the provided information by DSUE-DHI improves system efficiency by guiding some travelers to make better choices. However, as travelers' sensitivity to travel cost increases, the provided information exacerbates the competition among travelers. Namely, travelers tend to concentrate on the SPDT choices with lower travel costs, making traffic more congested.

Next, we explore the effects of the proportion of strategic travelers to see how strategic information transparency influences travelers' experience and system performance. Figure 16 illustrates the change of the relative difference between the informed and realized travel time. We can find that as the proportion of strategic class increases, the information accuracy decreases for both classes, while such decrease is slightly more severe for the regular class. And when there are no travelers receiving the strategic forecast information, the forecast information is identical to the realized travel time. These findings are in line with Proposition 4.

Figure 17 (a) shows the changes in average systematic disutility as the proportion of strategic travelers changes for DSUE-DHI, with DSUE as a reference (the dashed line). Note that, when the proportion equals to zero or one, there are no average systematic disutility that can be measured for the strategic class and the regular class respectively since there is no traveler in the corresponding class. Hence, we use approximations represented as dotted lines in the figure. Specifically, we use the cases where the proportion of strategic class equals 0.999 and 0.001 as approximations of the extreme cases where the proportion equals 1 and 0, respectively. The figure shows that, as the proportion of strategic travelers increases, the average systematic disutility of the whole increases slightly. However, there are sharper increases for each of the classes. This implies that the benefit gradually decreases as more travelers make decisions based on the strategic forecast information. Moreover, as the proportion of strategic travelers increases, regular travelers who did not switch to strategic class and the original strategic travelers feel less satisfied than before. However, the regular class travelers who switch to strategic class would feel more satisfied than before, though their perceived disutility may not be as decreased as expected. Hence, the disutility for the whole group does not vary as largely as each single class. When the proportion of strategic travelers is smaller than a threshold, around 0.8 in this example, strategic travelers at DSUE-DHI experience better (a lower disutility) than the DSUE case. However, regular travelers at DSUE-DHI always experience worse (a higher disutility) than the DSUE case.

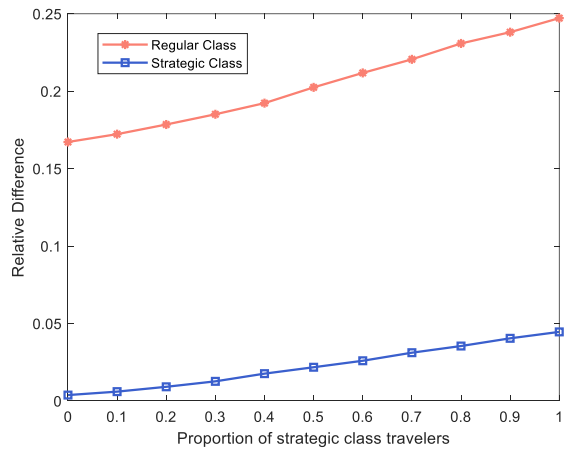

Figure 16. Relative difference between *ITT* and *RTT* as the proportion of strategic class changes

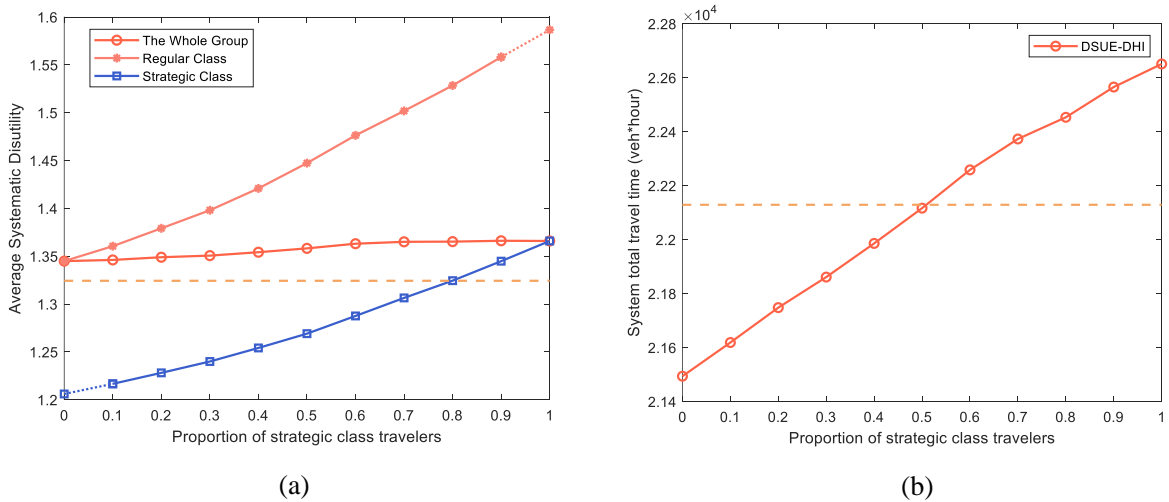

(a)                                 (b)

Figure 17. Changes of, (a) average systematic disutility, (b) system total travel time, as the proportion of strategic class changes at DSUE-DHI (dashed line: DSUE)

Figure 17 (b) shows at DSUE-DHI the total system travel time increases as the proportion of strategic travelers increases. And there is a threshold value of the proportion of strategic travelers, around 50%, smaller than which the system performance at DSUE-DHI could be better than the DSUE case. The results demonstrate that providing imperfect information, i.e., with the DSUE-DHI, can positively help congestion reduction. However, increasing the forecast information transparency may reduce the system performance.

Now, we assume travelers can choose to subscribe to either of the two types of information in the market. From the above results, the strategic class can feel more superior than the other class because of the larger gap between the two classes in terms of disutility as the size of strategic class increases. But the overall system performance and the general experiences (average disutility) of the whole group become worse. This implies that the superiority of the strategic class over the regular class becomes more significant when more travelers subscribe to the forecast information. Such superiority may attract more regular travelers to subscribe to the forecast information. However, the general experience of travelers and the system performance will worsen as the forecast information penetration increases. If a private company seeks profits by providing forecast information, it may attempt to attract as many subscribers as possible. This attempt, however, will increase congestion.

*7.3 DSUE-DHI with higher demand*

This section presents a higher-demand scenario where we doubled the demand compared to the base case. In the base case, the path travel time at the most congested departure time interval is 2.7 times of the free-flow travel time on average. And this number increases to 5.8 in the higher-demand scenario.

Figure 18 (a) and (b) show the relative difference between informed travel time and realized travel time as the value of $\theta$ and proportion of strategic changes, respectively. Compared with Figure 14 and Figure 16, Figure 18 (a) and (b) show similar trends as the congestion level increases. However, the information accuracy reduces when congestion increases, more significantly for the regular class. The superiority of strategic information in terms of information accuracy becomes larger as congestion increases. This implies that the traffic prediction of strategic information becomes more beneficial for a more congested network. In addition, when demand increases, the benefit of forecast information accuracy becomes salient even when $\theta$ is small. It implies, when the network is more congested, the superiority of strategic information in terms of accuracy remains significant even when travelers are less sensitive to travel costs.

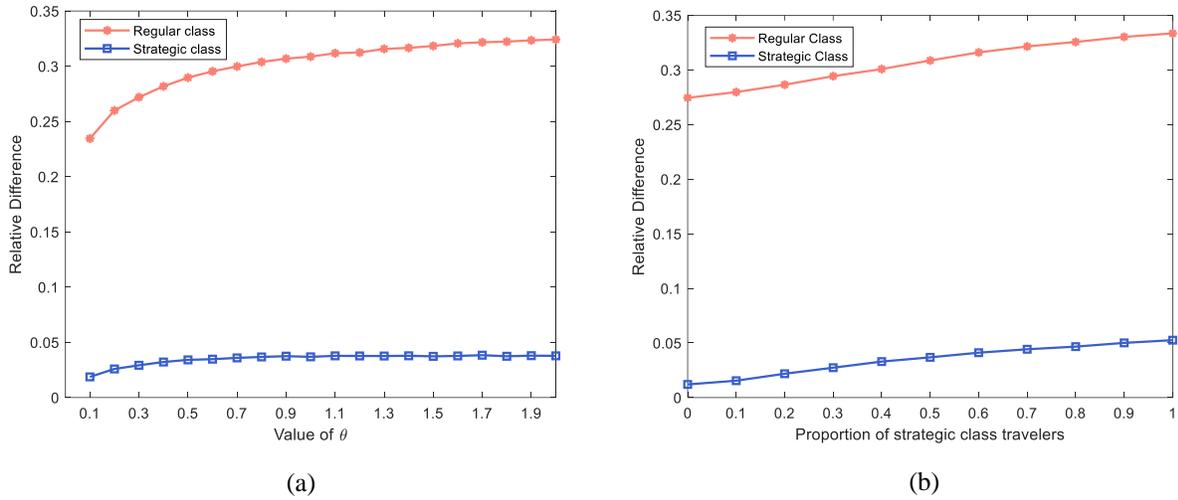

(a)                            (b)

Figure 18. Relative difference between *ITT* and *RTT* as, (a) value of $\theta$, (b) proportion of strategic class, changes

Figure 19 shows the systematic disutility variation as the value of $\theta$ changes for DSUE-DHI in the higher-demand scenario. We can see the increased average disutility induced by the increased demand. And, as $\theta$ increases, the average systematic disutility decreases, which is consistent with the lower demand scenario as shown in Figure 15 (a). However, the difference between the two classes and the difference between DSUE-DHI and DSUE have become more significant. This again implies that the traffic dynamics prediction in strategic information becomes increasingly beneficial for congested networks. Also, high information accuracy is more beneficial for congested networks. This leads to the lower average disutility at DSUE, in which travelers are assumed to perceive travel costs based on the realized travel times, compared with the strategic class when the value of $\theta$ is larger than 1.3.

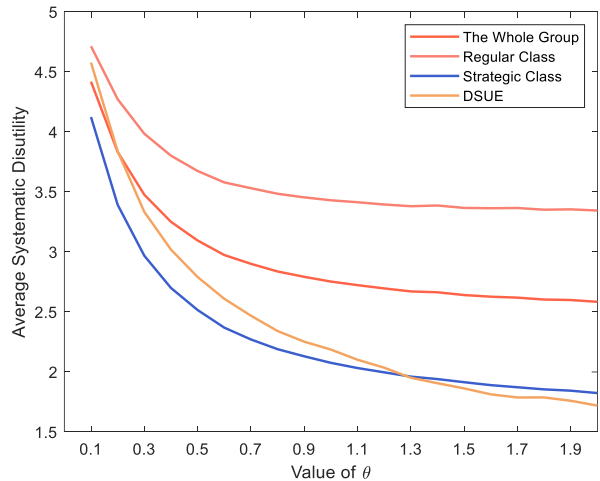

Figure 19. Average systematic disutility as the value of $\theta$ changes in the higher-demand scenario

*7.4 Discussion on solution convergence*

This section shows the computational convergence results with difference values of dispersion parameter

$\theta$ and the proportion of strategic travelers. The value defined in stopping condition (ii) in Step 3 of the algorithm in Section 6 is used as the convergence criterion. We also investigate the effect of the initial path departure pattern on the resulting equilibrium.

Figure 20 (a) shows the change in the value of convergence criterion as the algorithm iterates with different values of the dispersion parameter. It demonstrates convergence in all scenarios. In addition, when $\theta$ is smaller, the algorithm can converge faster. Note that the dispersion parameter affects network congestion. A larger $\theta$ means higher sensitivity to travel costs for travelers and thus contributes to the network's congestion. The result implies that when the congestion is minor, the algorithm can converge faster and smoother to the solution; however, when the congestion is significant, it needs more iterations to converge.

Figure 20 (b) shows the change of the convergence criterion under different proportions of the strategic travelers. When the proportion is smaller, the algorithm can converge faster. In line with previous analyses, the increased proportion of strategic travelers has a negative impact on the network performance, i.e., increasing network congestion, causing a slow speed of convergence. Note that the plots are in logarithmic scale.

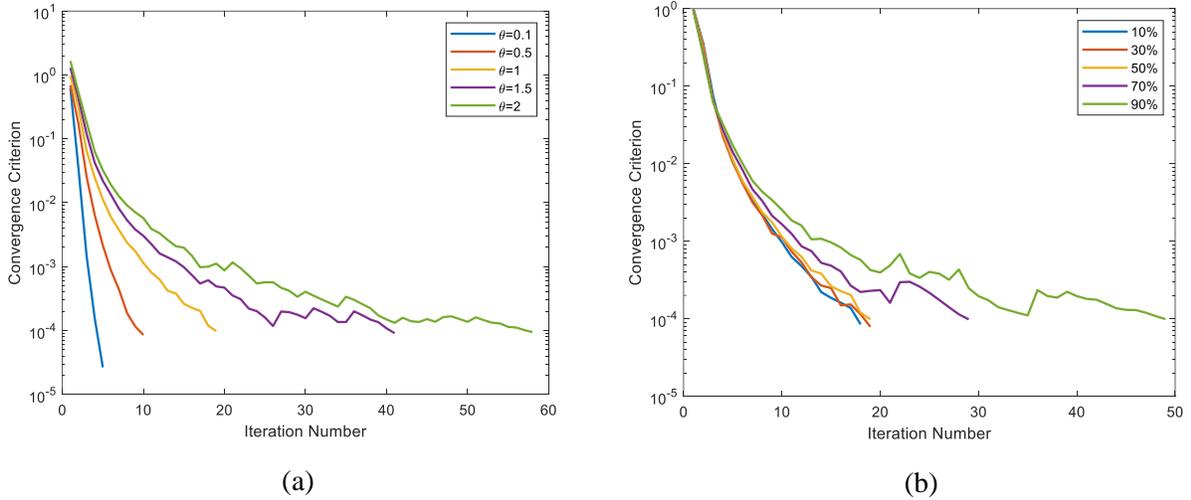

Figure 20. Convergence results with different values of (a) $\theta$ and (b) proportion of strategic class

Next, we randomly generate 1000 initial path departure patterns as the input to the SRAM algorithm, trying to reveal any influences of initial path departure patterns on the equilibrium. Here, the relative distance between any path departure $\mathbf{h}$ and the base case result $\mathbf{h}^B$ is defined as $\|\mathbf{h} - \mathbf{h}^B\|^2 / \|\mathbf{h}^B\|^2$. Figure 21 plots the histogram of the 1000 relative distances between the result solved in Section 7.1 and the results obtained using the randomly generated initial path departures. It shows that all solutions are close to the previously solved equilibrium with the maximum relative distance less than $5e-4$, demonstrating that, mathematically, the solutions concentrate on a small neighborhood. And, from the perspective of physical meaning, these solutions may represent a unique solution of path departures.

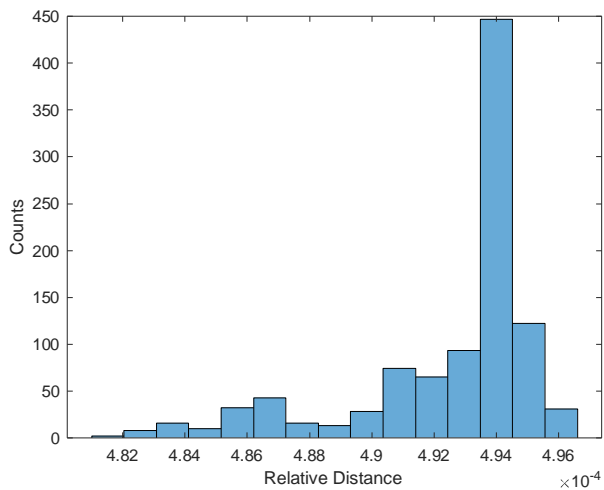

Figure 21. Distribution of relative distances to the base result

## 8. Conclusions

This paper develops a model for multi-class within-day dynamic traffic equilibrium, incorporating explicit information provision with two classes of travelers receiving regular instantaneous travel times and strategic forecast travel times, respectively. This explicit modeling approach accounts for information consistency and makes the information provision an endogenous element, differentiating itself from existing within-day dynamic equilibrium models. The established dynamic traffic equilibrium, labelled as Dynamic Stochastic User Equilibrium with Dynamic Heterogeneous Information (DSUE-DHI), is formulated as the solution to a fixed-point problem. Analytical properties of DSUE-DHI are presented and proved, with the impacts of travel time information further analyzed through numerical experiments on the Anaheim network.

The paper has shown that the classical DSUE is a special case of DSUE-DHI. The DSUE-DHI distinguishes itself from DSUE in congested scenarios; and it assimilates into DSUE in uncongested scenarios where instantaneous and strategic forecast travel times are identical to the realized travel times. Moreover, the proposed DSUE-DHI aligns closely with reality, acknowledging that travel time information can be accurate or perfect in uncongested scenarios but imperfect in congested scenarios. In congested scenarios, information provision in DSUE-DHI becomes an endogenous element critical in affecting network performance. Numerical results show that providing imperfect information, as in DSUE-DHI, can help improve system performance and travelers' experience compared to DSUE.

Theoretical results affirm that strategic forecast information is more accurate than regular instantaneous information; however, this accuracy reduces as more travelers receive the strategic forecast information. These properties are also validated through numerical experiments. In DSUE-DHI, travelers who receive strategic forecast information experience lower disutility and higher information reliability than those receiving instantaneous information. Yet, increasing the forecast information transparency may adversely affect system performance and the whole group's experiences in general on congested networks. The advantages of strategic forecast information become more pronounced and effective in reducing travelers' disutility when the network is more congested or when travelers are more sensitive to travel cost.

This research, in line with other studies, has also observed the *information paradox* (Yang and Jayakrishnan, 2013, Tavafoghi and Teneketzis, 2017, Acemoglu et al., 2018, Khan and Amin, 2018, Grzybowska et al., 2019, Jiang et al., 2020, Yu et al., 2020, Liu and Yang, 2021). Namely, more accurate information provided to more travelers can mitigate the benefits and even worsen system performance. When travelers have the option to subscribe to either type of information, the strategic forecast information would be appealing to travelers due to its accuracy and reliability. However, as the group size of travelers receiving forecast information increases, their travel cost increases, and system performance gets worse. This research underscores the need for careful operational design to mitigate the negative impact caused by information paradox.

This research could be further extended in several directions. First, while the proportions of the two classes of travelers are predetermined in this paper, future research could model the class composition of information adopters as a variable that stabilizes at equilibrium. Second, the present research assumes that strategic forecast information is free of charge in the modeling. This setting could be relaxed by introducing information pricing and fairness analysis. Third, operational control could be integrated into the information provision method to improve the traffic system performance, such as controls on the group sizes of different information receivers.

**Acknowledgments**

This work was supported in part by the U.S. National Science Foundation under Grant CMMI-2047793. The authors are solely responsible for the contents of this paper.

**Appendix A:**

This appendix presents the dynamic network loading (DNL) model used in the numerical examples of this paper. The DNL model maps from realized path departures to traffic condition, i.e., link/path travel time. Since the modeling of DNL is a distinct research topic and not the focus of this paper, we directly adopt the one proposed in Han et al. (2019) with a minor adaptation. Note that other DNL models can also be applied within the proposed modeling framework.

The DNL model proposed in Han et al. (2019) follows the first-in-first-out (FIFO) principle and is formulated as a differential-algebraic equation (DAE) system. It encompasses models of link dynamics, link demand and supply, junction dynamics, queueing, and link/path travel times. This paper makes the following modification to further ensure flow conservation.

The link demand function $D(t)$, i.e., Eq. (3.8) in Han et al. (2019), has been revised to $\min\left(C, \left(N^{up}\left(t - \frac{L}{v}\right) - N^{dn}(t)\right)/dt\right)$ instead of using capacity $C$ under the condition $N^{up}\left(t - \frac{L}{v}\right) > N^{dn}(t)$, where $N^{up}$ and $N^{dn}$ represent the upstream and downstream cumulative flows, $L$ the link length, $v$ the forward kinematic wave speed, and $dt$ the length of each time interval. The revised formula helps ensure the flow conservation more rigorously. Based on the FIFO principle, the link demand at time $t$ must not exceed the flow rate corresponding to the flow that has entered the link at time $t - \frac{L}{v}$ but has not exited the link at time $t$. This rate can be smaller than the link capacity. Hence, the link demand should be the smaller one between this rate and the capacity. Any flow that entered the link after time $t - \frac{L}{v}$ will not be counted into the link demand at time $t$ since such flow must have not reached the end of the link at time $t$. For other details about the DAE system, please refer to Han et al. (2019).

## Appendix B: Proof of Proposition 5.

**Proposition 5.** At equilibrium of DSUE-DHI, the accuracy of forecast information is higher than or equal to that of instantaneous information in the Euclidean norm, if the $DNL$ mapping is continuous and convex. Namely, the following inequality holds if the $DNL$ mapping is continuous and convex:

$$\|\mathbf{\Phi}_I(\mathbf{h}^*) - \bar{\mathbf{t}}(\mathbf{h}^*)\| \geq \|\mathbf{\Phi}_F(\mathbf{h}^*) - \bar{\mathbf{t}}(\mathbf{h}^*)\| \tag{17}$$

where $\bar{\mathbf{t}}$ contains the realized travel times for all pairs of departure time $t \in \mathcal{T}$ and path $p \in \mathcal{P}$. $\mathbf{\Phi}_I$ contains the instantaneous travel time information provided at each time $t \in \mathcal{T}$ for each path $p \in \mathcal{P}$, and $\mathbf{\Phi}_F$ contains the forecast travel time information provided at each time $t \in \mathcal{T}$ for departure time $t$ and each path $p \in \mathcal{P}$. $\mathbf{h}^*$ represents the path departures at DSUE-DHI.

**Proof.** For the sake of simplicity and convenience, any variables in matrix form defined in the paper are rearranged as column vectors by stacking all columns into one in this proof unless specifically noted.

To prove Eq. (17) is to prove the following function $F(\mathbf{h}^*) \geq 0, \forall \mathbf{h}^* \geq 0$,

$$F(\mathbf{h}^*) = \|\mathbf{\Phi}_I(\mathbf{h}^*) - \bar{\mathbf{t}}(\mathbf{h}^*)\|^2 - \|\mathbf{\Phi}_F(\mathbf{h}^*) - \bar{\mathbf{t}}(\mathbf{h}^*)\|^2.$$

The derivative of function $F$ with respect to $\mathbf{h}^*$ is

$$\frac{\partial F(\mathbf{h}^*)}{\partial \mathbf{h}^*} = 2 \left( \frac{\partial (\mathbf{\Phi}_I(\mathbf{h}^*) - \bar{\mathbf{t}}(\mathbf{h}^*))}{\partial \mathbf{h}^*} \right)^T (\mathbf{\Phi}_I(\mathbf{h}^*) - \bar{\mathbf{t}}(\mathbf{h}^*)) - 2 \left( \frac{\partial (\mathbf{\Phi}_F(\mathbf{h}^*) - \bar{\mathbf{t}}(\mathbf{h}^*))}{\partial \mathbf{h}^*} \right)^T (\mathbf{\Phi}_F(\mathbf{h}^*) - \bar{\mathbf{t}}(\mathbf{h}^*)) \tag{18}$$

Let $\widehat{DNL}$ be a $(P \times T \times T) \times 1$ vector consisting of $DNL(\mathbf{h}_{\tilde{t}}^*)$, $\tilde{t} \in \mathcal{T}$, with a vertical arrangement from $t_1$ to $t_T$, where $P$ equals the number of paths. Then,

$$\mathbf{\Phi}_F(\mathbf{h}^*) = W \cdot \widehat{DNL},$$

where $W$ is a $(P \times T) \times (P \times T \times T)$ matrix and

$$W(i,j) = \begin{cases} 1, & if\ j = \lfloor i/P \rfloor \times P \times (T+1) + (i - \lfloor i/P \rfloor \times P) \\ 0, & o.w. \end{cases}.$$

Then,

$$\frac{\partial \mathbf{\Phi}_F(\mathbf{h}^*)}{\partial \mathbf{h}^*} = W \frac{\partial \widehat{DNL}}{\partial \mathbf{h}^*}.$$

Let $Z^{\tilde{t}}, \tilde{t} \in \mathcal{T}$ be a $(P \times T \times T) \times (P \times T)$ matrix where the $L(\tilde{t})$-th $P \times T$ rows compose an identity matrix with dimension $(P \times T) \times (P \times T)$ and anywhere else is zero. $L(\tilde{t})$ represents the ordinal number of the time interval $\tilde{t}$. Then we have

$$\widehat{DNL} = \sum_{\tilde{t}} Z^{\tilde{t}} \cdot DNL(\mathbf{h}_{\tilde{t}}^*).$$

Let $A^{\tilde{t}}, \tilde{t} \in \mathcal{T}$ be a $(P \times T) \times (P \times T)$ matrix where the first $P \times (L(\tilde{t}) - 1)$ rows and first $P \times (L(\tilde{t}) - 1)$ columns compose an identity matrix and anywhere else is zero. Let $B^{\tilde{t}}, \tilde{t} \in \mathcal{T}$ be a $(P \times T) \times (P \times (T - L(\tilde{t}) + 1))$ matrix where the last $P \times (T - L(\tilde{t}) + 1)$ rows compose an identity matrix, and anywhere else is zero. Then we have,

$$\boldsymbol{h}^*_{\tilde{t}} = A^{\tilde{t}} \cdot \mathbf{h}^* + B^{\tilde{t}} \cdot \bar{\mathbf{h}}^*_{\tilde{t}}.$$

Therefore,

$$\begin{aligned}
\frac{\partial \boldsymbol{\Phi}_F}{\partial \mathbf{h}^*} &= W \cdot \frac{\partial \widetilde{DNL}}{\partial \mathbf{h}^*} \\
&= W \cdot \sum_{\tilde{t}} \left( Z^{\tilde{t}} \cdot \frac{\partial DNL(\boldsymbol{h}^*_{\tilde{t}})}{\partial \mathbf{h}^*} \right) \\
&= W \cdot \sum_{\tilde{t}} \left( Z^{\tilde{t}} \cdot \frac{\partial DNL(\boldsymbol{h}^*_{\tilde{t}})}{\partial \boldsymbol{h}^*_{\tilde{t}}} \cdot \frac{\partial \boldsymbol{h}^*_{\tilde{t}}}{\partial \mathbf{h}^*} \right) \\
&= W \cdot \sum_{\tilde{t}} \left( Z^{\tilde{t}} \cdot \frac{\partial DNL(\boldsymbol{h}^*_{\tilde{t}})}{\partial \boldsymbol{h}^*_{\tilde{t}}} \cdot \left( A^{\tilde{t}} + B^{\tilde{t}} \cdot \frac{\partial \bar{\mathbf{h}}^*_{\tilde{t}}}{\partial \mathbf{h}^*} \right) \right) \\
&= \sum_{\tilde{t}} \left( W \cdot Z^{\tilde{t}} \cdot \frac{\partial DNL(\boldsymbol{h}^*_{\tilde{t}})}{\partial \boldsymbol{h}^*_{\tilde{t}}} \cdot A^{\tilde{t}} + W \cdot Z^{\tilde{t}} \cdot \frac{\partial DNL(\boldsymbol{h}^*_{\tilde{t}})}{\partial \boldsymbol{h}^*_{\tilde{t}}} \cdot B^{\tilde{t}} \cdot \frac{\partial \bar{\mathbf{h}}^*_{\tilde{t}}}{\partial \mathbf{h}^*} \right).
\end{aligned}$$

Since $W \cdot Z^{\tilde{t}} \cdot \frac{\partial DNL(\boldsymbol{h}^*_{\tilde{t}})}{\partial \boldsymbol{h}^*_{\tilde{t}}} \cdot A^{\tilde{t}} = \mathbf{0}, \forall \tilde{t} \in \mathcal{T}$, no matter what $\frac{\partial DNL(\boldsymbol{h}^*_{\tilde{t}})}{\partial \boldsymbol{h}^*_{\tilde{t}}}$ equals to, then

$$\frac{\partial \boldsymbol{\Phi}_F}{\partial \mathbf{h}^*} = \sum_{\tilde{t}} \left( W \cdot Z^{\tilde{t}} \cdot \frac{\partial DNL(\boldsymbol{h}^*_{\tilde{t}})}{\partial \boldsymbol{h}^*_{\tilde{t}}} \cdot B^{\tilde{t}} \cdot \frac{\partial \bar{\mathbf{h}}^*_{\tilde{t}}}{\partial \mathbf{h}^*} \right).$$

According to the definitions of $W$, $Z^{\tilde{t}}$ and $\boldsymbol{h}^*_{\tilde{t}}$, we have

$$\begin{aligned}
\frac{\partial \boldsymbol{\Phi}_F}{\partial \mathbf{h}^*} &= \sum_{\tilde{t}} \left( W \cdot Z^{\tilde{t}} \cdot \sum_{\tilde{t}} \left( W \cdot Z^{\tilde{t}} \cdot \frac{\partial DNL(\boldsymbol{h}^*_{\tilde{t}})}{\partial \boldsymbol{h}^*_{\tilde{t}}} \right) \cdot B^{\tilde{t}} \cdot \frac{\partial \bar{\mathbf{h}}^*_{\tilde{t}}}{\partial \mathbf{h}^*} \right) \\
&= \sum_{\tilde{t}} \left( W \cdot Z^{\tilde{t}} \cdot \frac{\partial DNL(\mathbf{h}^*)}{\partial \mathbf{h}^*} \cdot B^{\tilde{t}} \cdot \frac{\partial \bar{\mathbf{h}}^*_{\tilde{t}}}{\partial \mathbf{h}^*} \right) \\
&= \frac{\partial DNL(\mathbf{h}^*)}{\partial \mathbf{h}^*} \cdot \sum_{\tilde{t}} \left( W \cdot Z^{\tilde{t}} \cdot B^{\tilde{t}} \cdot \frac{\partial \bar{\mathbf{h}}^*_{\tilde{t}}}{\partial \mathbf{h}^*} \right).
\end{aligned} \quad (19)$$

At the DSUE-DHI, we have

$$\lambda \sum_{\tilde{t}} (W \cdot Z^{\tilde{t}} \cdot B^{\tilde{t}} \cdot \bar{\mathbf{h}}^*_{\tilde{t}}) = \mathbf{h}^*_I,$$

where $\lambda \in [0,1]$ is the proportion of the travelers who receive the instantaneous information.

Hence, we have

$$\lambda \sum_{\tilde{t}} \left( W \cdot Z^{\tilde{t}} \cdot B^{\tilde{t}} \cdot \frac{\partial \bar{\mathbf{h}}^*_{\tilde{t}}}{\partial \mathbf{h}^*} \right) = \frac{\partial \mathbf{h}^*_I}{\partial \mathbf{h}^*},$$

and then

$$\sum_{\tilde{t}} \left( W \cdot Z^{\tilde{t}} \cdot B^{\tilde{t}} \cdot \frac{\partial \bar{\mathbf{h}}_{\tilde{t}}^*}{\partial \mathbf{h}^*} \right) = \frac{1}{\lambda} \cdot \frac{\partial \mathbf{h}_I^*}{\partial \mathbf{h}^*} = \frac{1}{\lambda} \cdot \lambda \cdot I = I, \tag{20}$$

where $I$ is an identity matrix with dimension $(P \times T) \times (P \times T)$.

Insert Eq. (20) into Eq. (19) and we can obtain

$$\frac{\partial \mathbf{\Phi}_F(\mathbf{h}^*)}{\partial \mathbf{h}^*} = \frac{\partial DNL(\mathbf{h}^*)}{\partial \mathbf{h}^*} \cdot I = \frac{\partial \bar{\mathbf{t}}(\mathbf{h}^*)}{\partial \mathbf{h}^*}. \tag{21}$$

Then, insert Eq. (21) into Eq. (18) and we have

$$\frac{\partial F(\mathbf{h}^*)}{\partial \mathbf{h}^*} = 2 \left( \frac{\partial \mathbf{\Phi}_I(\mathbf{h}^*)}{\partial \mathbf{h}^*} - \frac{\partial \bar{\mathbf{t}}(\mathbf{h}^*)}{\partial \mathbf{h}^*} \right)^T \left( \mathbf{\Phi}_I(\mathbf{h}^*) - \bar{\mathbf{t}}(\mathbf{h}^*) \right). \tag{22}$$

If the mapping $DNL$ is continuous, there exists a path departure vector $\mathbf{h}'$ that satisfies $\bar{\mathbf{t}}(\mathbf{h}') = \mathbf{\Phi}_I(\mathbf{h}^*)$. Since both $\mathbf{\Phi}_I(\mathbf{h}^*)$ and $\bar{\mathbf{t}}(\mathbf{h}')$ are directly obtained through the DNL model, given that $\bar{\mathbf{t}}(\mathbf{h}') = \mathbf{\Phi}_I(\mathbf{h}^*)$, $\mathbf{\Phi}_I(\mathbf{h}^* + \Delta \mathbf{h}^*) \approx \bar{\mathbf{t}}(\mathbf{h}' + \Delta \mathbf{h}')$ holds where $\Delta \mathbf{h}^*$ and $\Delta \mathbf{h}'$ are infinitesimal increments with respect to $\mathbf{h}^*$ and $\mathbf{h}'$. That is, $\frac{\partial \mathbf{\Phi}_I(\mathbf{h}^*)}{\partial \mathbf{h}^*} \approx \frac{\partial \bar{\mathbf{t}}(\mathbf{h}')}{\partial \mathbf{h}'}$ holds. Then Eq. (22) can be written as

$$\frac{\partial F(\mathbf{h}^*)}{\partial \mathbf{h}^*} = 2 \left( \frac{\partial \bar{\mathbf{t}}(\mathbf{h}')}{\partial \mathbf{h}'} - \frac{\partial \bar{\mathbf{t}}(\mathbf{h}^*)}{\partial \mathbf{h}^*} \right)^T \left( \bar{\mathbf{t}}(\mathbf{h}') - \bar{\mathbf{t}}(\mathbf{h}^*) \right).$$

If the mapping $DNL$ is convex, the following inequality holds for any $\mathbf{h}^*, \mathbf{h}' \geq \mathbf{0}$.

$$\left( \frac{\partial \bar{\mathbf{t}}(\mathbf{h}')}{\partial \mathbf{h}'} - \frac{\partial \bar{\mathbf{t}}(\mathbf{h}^*)}{\partial \mathbf{h}^*} \right)^T \left( \bar{\mathbf{t}}(\mathbf{h}') - \bar{\mathbf{t}}(\mathbf{h}^*) \right) \geq \mathbf{0}$$

That is, $F$ is non-decreasing:

$$\frac{\partial F(\mathbf{h}^*)}{\partial \mathbf{h}^*} \geq \mathbf{0}.$$

When $\mathbf{h}^* = \mathbf{0}$, both the instantaneous travel time and forecast travel time equal to the realized travel time (free-flow travel time), i.e., $F(\mathbf{h}^*) = 0$. Therefore, given that $F$ is non-decreasing, we have $F(\mathbf{h}^*) \geq 0, \forall \mathbf{h}^* \geq \mathbf{0}$. The proof is complete. ∎